\documentclass[journal,comsoc]{IEEEtran}

\usepackage{xpatch}
\newcounter{mybibstartvalue}
\usepackage{etoolbox}
\usepackage{balance}
\usepackage[T1]{fontenc}
\usepackage{cite}
\usepackage{amsmath}
\usepackage{amsmath,amssymb,amsfonts}
\usepackage{graphicx}
\usepackage{textcomp}
\usepackage{graphicx}
\usepackage{textcomp}
\usepackage[cmintegrals]{newtxmath}
\usepackage{blindtext, graphicx}
\usepackage{enumitem}
\usepackage{colortbl,hhline}
\usepackage[binary-units=true]{siunitx}
\usepackage[table]{xcolor}
\usepackage{algpseudocode}
\usepackage{algorithm}
\usepackage{amssymb}
\usepackage{array}
\usepackage{soul}
\ifCLASSOPTIONcompsoc
\usepackage[caption=false,font=footnotesize,labelfont=sf,textfont=sf]{subfig}
\else
\usepackage[caption=false,font=footnotesize]{subfig}
\fi

\usepackage{dblfloatfix}
\usepackage{booktabs}
\usepackage{multirow}
\usepackage{tabularx}
\usepackage{adjustbox}
\usepackage{makecell}
\usepackage{threeparttable}
\usepackage{algpseudocode}
\usepackage{soul}
\usepackage{float}
\algnewcommand{\algorithmicand}{\textbf{ and }}
\algnewcommand{\algorithmicor}{\textbf{ or }}
\algnewcommand{\OR}{\algorithmicor}
\algnewcommand{\AND}{\algorithmicand}
\usepackage[font=footnotesize]{caption}
\usepackage{etoolbox}

\algnewcommand\algorithmicforeach{\textbf{for each}}
\algdef{S}[FOR]{ForEach}[1]{\algorithmicforeach\ #1\ \algorithmicdo}

\newcommand{\algbrakline}[1]{\newline\makebox[#1cm]{}}
\newcommand{\tableRef}[1]{Table~\ref{#1}}
\newcommand{\figureRef}[1]{Fig.~\ref{#1}}
\newcommand{\sectionRef}[1]{Section~\ref{#1}}

\newcommand{\behnaz}[1]{\textcolor{black}{#1}}
\definecolor{BrickRed}{RGB}{255,0,40}

\newcommand{\answersnd}[1]{\textcolor{black}{#1}}

\usepackage[all]{background}
\usepackage{lipsum}
\SetBgContents{\textcolor{blue}{This is an extended version of the paper accepted to be published in IEEE (TCAD) 2021 which includes Appendices A-E}}
\SetBgPosition{-0.8cm,-9.0cm}
\SetBgOpacity{1.0}
\SetBgAngle{90.0}
\SetBgScale{1.3}

\begin{document}



\title{Toward the Design of Fault-Tolerance- and Peak- Power-Aware Multi-Core Mixed-Criticality Systems}
%
%

\author{Behnaz~Ranjbar,
       Ali~Hosseinghorban,
       Mohammad~Salehi,
       Alireza~Ejlali,
       and~Akash~Kumar,~\IEEEmembership{Senior Member, IEEE}
\thanks{Manuscript received September 30, 2020; revised December 28, 2020 and April 8, 2021, accepted in May 13, 2021. This article was recommended by Associate Editor Subhasish Mitra. (Corresponding author: Akash Kumar)}    
\thanks{Behnaz Ranjbar and Ali Hosseinghorban are with the Chair for Processor Design, CFAED, Technische Universit{\"a}t Dresden, 01062 Dresden, Germany, and also with the Department of Computer Engineering, Sharif University of Technology, Tehran 11365-11155, Iran. e-mail:~behnaz.ranjbar@tu-dresden.de, ali.hosseinghorban@mailbox.tu-dresden.de}
\thanks{Mohammad Salehi is with the Department of Computer Engineering, University of Guilan, Rasht, Iran. e-mail:~mohammad.salehi@guilan.ac.ir}
\thanks{Alireza Ejlali is with the Department of Computer Engineering, Sharif University of Technology, Tehran 11365-11155, Iran. e-mail:~ejlali@sharif.edu}
\thanks{Akash Kumar is with the Chair for Processor Design, CFAED, Technische Universit{\"a}t Dresden, 01062 Dresden, Germany. e-mail:~akash.kumar@tu-dresden.de}
\thanks{Digital Object Identifier: 10.1109/TCAD.2021.3082495}
}

%
%

\markboth{IEEE Transactions on Computer-Aided Design of Integrated Circuits and Systems
}%
{Ranjbar \MakeLowercase{\textit{et al.}}: Toward the Design of Peak Power- and Fault-Tolerance-Aware Multi-core Mixed-Criticality Systems}
%




\maketitle

\begin{abstract}

Mixed-Criticality (MC) systems have recently been devised to address the requirements of real-time systems in industrial applications, where the system runs tasks with different criticality levels on a single platform.
In some workloads, a high-critically task might overrun and overload the system, or a fault can occur during the execution.
However, these systems must be fault-tolerant and guarantee the correct execution of all high-criticality tasks by their deadlines to avoid catastrophic consequences, in any situation.
Furthermore, in these MC systems, the peak power consumption of the system may increase, especially in an overload situation and exceed the processor Thermal Design Power (TDP) constraint. This may cause generating heat beyond the cooling capacity, resulting the system stop to avoid excessive heat and halting the processor.
In this paper, we propose a technique for dependent dual-criticality tasks in fault-tolerant multi-core MC systems to manage peak power consumption and temperature. 
The technique develops a tree of possible task mapping and scheduling at design-time to cover all possible scenarios and reduce the low-criticality task drop rate in the high-criticality mode.
At run-time, the system exploits the tree to select a proper schedule according to fault occurrences and criticality mode changes. 
Experimental results show that the average task schedulability is 74.14\% on average for the proposed method, while the peak power consumption and maximum temperature are improved by 16.65\% and 14.9$^\circ$C on average, respectively, compared to a recent work.
In addition, for a real-life application, our method reduces the peak power and maximum temperature by up to 20.06\% and 5$^\circ$C, respectively, compared to a state-of-the-art approach.

\end{abstract}


%
\IEEEpeerreviewmaketitle

\section{Introduction}
\label{sec:introduction}

\IEEEPARstart{R}{ecently}, multi-core real-time embedded systems where industrial applications with real-time constraints are executed, becomes significant. 
These systems are responsible for executing an application with a set of High-Criticality (HC) and Low-Criticality (LC) tasks on a single platform~\cite{burns13,Ranjbar2020Fantom,Bahrami2021}. 
For instance, aerospace applications have flight-critical tasks that must be executed before the deadline and mission-oriented tasks that are important, but they could be dropped in critical situations~\cite{Ranjbar2020Fantom}. 
These systems are called Mixed-Criticality (MC) systems.

In order to map and schedule these MC tasks on multi-core platforms, the Worst-Case Execution Time (WCET) of each task is exploited to guarantee the correct execution of all tasks, especially HC tasks, in any situation.
However, the WCET of a task is pessimistic, and the probability that the execution time of the task will be as large as WCET is very low~\cite{Ernst2016}.
Therefore, most of the time, the system processing capacity is wasted because the execution time of tasks is less than the pessimistic WCETs.
To this end, MC systems consider two or multiple WCETs for HC tasks~\cite{burns13, Ranjbar2020, Ernst2016}, which are calculated using different WCET-determination methodologies and tools (e.g., calculating pessimistic or optimistic WCET values). 
In this paper, we consider a dual-criticality system where each task has a low WCET and a high WCET.

\behnaz{In MC systems, if an HC task overruns (the task's execution time exceeds the low WCET), the system switches to a high-criticality mode (HI mode). 
In this mode, the system considers the high WCET for all the remaining tasks to guarantee the safety of the system, and it stays in this mode until there is no HC task in the ready queue of each core~\cite{burns13,Ranjbar2020Fantom,Bahrami2021}.
In this situation, the execution of all LC and HC tasks requires higher computational demands, which may exceed the processor's capacity, and the system becomes overloaded~\cite{anderson09}. 
Thus, all cores may execute tasks simultaneously to meet deadlines of tasks, which increase the instantaneous processor power beyond its Thermal Design Power (TDP) constraint~\cite{lee14, munawar14, lee10}.} 

\answersnd{TDP is the maximum sustainable power that a chip can dissipate safely. 
The chip's power consumption is the sum of power consumption of all cores regardless of which kind of tasks (LC or HC) they are executing. So, even underestimating the power consumption of an LC task might violate the TDP constraint
which generates a large amount of heat that is higher than the cooling capacity of the chip. 
Thus, it may stop or restart the system by an integrated Dynamic Thermal Management (DTM) unit~\cite{munawar14}. 
Consequently, the deadlines of HC tasks are missed, which leads to catastrophic damages. 
Therefore, it is necessary to consider pessimistic power consumption value for all tasks with any criticality level.
Note that, minimizing only the average power consumption is not sufficient. Although it may decrease the instantaneous power, there is no guarantee that the TDP is not violated~\cite{munawar14}.}



In addition to meeting deadlines in case of tasks overrun, MC systems must use fault-tolerance techniques like task re-execution or duplication~\cite{koren07} to guarantee correct functionality when tasks are affected by transient faults~\cite{Ranjbar2020Fantom,sahoo2021reliability}. 
It should be mentioned that it is possible that the system faces both task overrun and one or multiple transient faults in one run. 
So, the system must be designed carefully to execute all HC tasks correctly before their deadlines, even in worst-case situations.

Previous works have proposed methods to schedule MC tasks in both low-criticality mode (LO mode) and HI mode; however, most of them have considered independent periodic MC tasks. 
These methods are not suitable for MC tasks with precedence constraints. 
Besides, some research works have proposed methods to reduce average power consumption in MC systems~\cite{huang14, guo14, ali15, narayana16, taherin15,Awan2016}. 
These works use the Dynamic Voltage and Frequency Scaling (DVFS) technique~\cite{hsu03} along with dropping LC tasks in the HI mode to manage average power consumption; however, none of them has considered managing peak power consumption. 
\behnaz{It is noteworthy to mention that some low-power techniques like DVFS, cannot be easily used in the HI mode, specially in the overload situations~\cite{huang14, guo14, ali15}, because changing the voltage and frequency levels of cores imposes high switching time overhead~\cite{lee14} that may cause deadline violation of HC tasks and also degrade the reliability level of tasks.}
On the other hand, some studies addressed peak power management in hard real-time systems with only one criticality level; therefore, they are not suitable for MC systems~\cite{lee14, munawar14, lee10}.
\tableRef{Table: related} presents a comparison between the state-of-the-art approaches which we discuss in detail in Section~\ref{sec:relatedWork}.

In this paper, we consider fault-tolerant MC multi-core embedded systems with low and high criticality tasks. 
The proposed method manages peak power and temperature to prevent hot spots in homogeneous multi-core platforms.
The task set is characterized by a directed acyclic graph (DAG)~\cite{baruah16}, and faults are tolerated through task re-execution. 
Since tasks can overrun and a fault can occur at any time but occasionally, using a single task's mapping and scheduling to guarantee the correct and on-time execution of all HC tasks without TDP violation leads to inefficient utilization of resources.

To this end, we propose a method that exploits a tree of schedules for dependent dual-criticality tasks. 
The proposed technique generates a tree of schedules off-line (at design-time) considering all possibility of fault occurrence scenarios in different tasks (including both LC and HC tasks) and HC task overrun.
At run-time, when an HC task overruns or a fault occurs in an LC or HC task, the scheduler chooses the proper schedule from the tree to tolerate the faults or manage the system mode switches with low overheads. 
Moreover, typical MC systems drop or degrade most of the LC tasks in the HI mode to guarantee the execution of all HC tasks. 
Therefore, another goal of our technique is to improve the LC tasks' QoS (Quality of Service, the percentage of executed LC tasks to all LC tasks~\cite{li18,Ranjbar2020Fantom}) in the HI mode while all HC tasks meet their deadlines. 
As a result, by generating the schedule tree and exploiting it at run-time, the LC tasks' QoS is maximized, while the peak power consumption of the system is managed and also the occurrence of possible faults is tolerated.

\textit{\textbf{Contributions:}} To the best of our knowledge, this paper is the first work to study the scheduling problem for fault-tolerant MC systems with peak power and thermal consideration. 
The main contributions of this paper are:

\begin{itemize}
	\item Proposing a tree generation approach for mixed-criticality systems, based on the all possibility of fault occurrence scenarios and criticality mode changes. 
	\item Peak power-aware task mapping and scheduling in multi-core mixed-criticality systems for both LO and HI modes.
	\item Offline QoS-Aware task mapping and scheduling to guarantee the correct execution of most LC tasks in the HI mode.
	\item Reducing the run-time timing overheads by generating all schedules at design time and exploiting them at run-time.
\end{itemize}

To evaluate our proposed method, we obtain the peak power consumption of tasks by running the benchmarks on the ARM Cortex-A7 core of the ODROID XU3 platform. Then, with the help of the platform's extracted information, we evaluate our method and state-of-the-art methods using HOTSPOT~\cite{huang06} simulator.
\behnaz{The experiments show that our scheme can schedule 74.14\% of task sets on average (43.04\% more, compared to~\cite{Medina18}), in which the TDP constraint and all deadlines are met, while the system can tolerate occurrence of up to a certain number of faults (four in our experiments). For a real task graph, our method reduces the peak power and maximum temperature by up to 20.06\% and 5$^\circ$C 
respectively, compared to the approach of~\cite{Medina18}, while the QoS is improved 9.09\%. 
On the other hand, although our method increases the maximum temperature by 9.61\%, compared to~\cite{socci15}, we reduce the peak power by 6.31\% and improve the QoS by 81.82\%.}

The rest of the paper is organized as follows. In Section~\ref{sec:relatedWork}, we review related works. In Section~\ref{sec:models}, we introduce models, assumptions and define the problem. The motivational example is presented in Section~\ref{sec:proposalMethod} and then we describe our method in detail. Finally, we analyze the experiments and conclude the paper in Sections~\ref{sec:expriment} and~\ref{sec:conclusion}, respectively.
 
\section{Related Work}
\label{sec:relatedWork}

\begin{table*}[t]
	\caption{Summary of state-of-the-art approaches}
	\label{Table: related}
	\footnotesize
	\centering
    \adjustbox{max width=\textwidth}{
	\begin{tabular}{ccccccccc}
	\hline 
	
		\toprule
		\multicolumn{1}{c}{\centering \#} &
		\multicolumn{1}{c}{\centering } &
		\multicolumn{1}{c}{\centering TDP Manag.} &
		\multicolumn{1}{c}{\centering Avg. Power} &
		\multicolumn{1}{c}{\centering Temp.} &
		\multicolumn{1}{c}{\centering Fault-Tolerance} &
		\multicolumn{1}{c}{\centering MC Tasks} &
		\multicolumn{1}{c}{\centering DAG Model} &
		\multicolumn{1}{c}{\centering LC tasks' QoS}\\

		\hline
		\hline
		
		\multicolumn{1}{c}{\centering 1} &
		\multicolumn{1}{c}{\centering \makecell{ Socci'15\cite{socci15},~Socci'19\cite{Socci19}, \\ Baruah'16\cite{baruah16}, Medina'17\cite{Medina17}}} &
		\multicolumn{1}{c}{\centering \textcolor{red}{$\times$}} &
		\multicolumn{1}{c}{\centering \textcolor{red}{$\times$}} &
		\multicolumn{1}{c}{\centering \textcolor{red}{$\times$}} &
		\multicolumn{1}{c}{\centering \textcolor{red}{$\times$}} &
		\multicolumn{1}{c}{\centering \textcolor{blue}{$\checkmark$}} &
		\multicolumn{1}{c}{\centering \textcolor{blue}{$\checkmark$}} &
		\multicolumn{1}{c}{\centering \textcolor{red}{$\times$}}\\

		\hline
		\multicolumn{1}{c}{\centering 2} &
		\multicolumn{1}{c}{\centering Li'16\cite{Li16},~Pathan'18\cite{pathan18}} &
		\multicolumn{1}{c}{\centering \textcolor{red}{$\times$}} &
		\multicolumn{1}{c}{\centering \textcolor{red}{$\times$}} &
		\multicolumn{1}{c}{\centering \textcolor{red}{$\times$}} &
		\multicolumn{1}{c}{\centering \textcolor{red}{$\times$}} &
		\multicolumn{1}{c}{\centering \textcolor{blue}{$\checkmark$}} &
		\multicolumn{1}{c}{\centering \textcolor{blue}{$\checkmark$}} &
		\multicolumn{1}{c}{\centering \textcolor{blue}{$\checkmark$}}\\
		
		\hline
		\multicolumn{1}{c}{\centering \cellcolor[HTML]{E3E2E2}3} &
		\multicolumn{1}{c}{\centering \cellcolor[HTML]{E3E2E2} Medina'18\cite{Medina18,Medina181},~Choi'18\cite{Choi18},~Bolchini'13\cite{Bolchini13}} &
		\multicolumn{1}{c}{\centering \cellcolor[HTML]{E3E2E2}\textcolor{red}{$\times$}} &
		\multicolumn{1}{c}{\centering \cellcolor[HTML]{E3E2E2}\textcolor{red}{$\times$}} &
		\multicolumn{1}{c}{\centering \cellcolor[HTML]{E3E2E2}\textcolor{red}{$\times$}} &
		\multicolumn{1}{c}{\centering \cellcolor[HTML]{E3E2E2}\textcolor{blue}{$\checkmark$}} &
		\multicolumn{1}{c}{\centering \cellcolor[HTML]{E3E2E2}\textcolor{blue}{$\checkmark$}} &
		\multicolumn{1}{c}{\centering \cellcolor[HTML]{E3E2E2}\textcolor{blue}{$\checkmark$}} &
		\multicolumn{1}{c}{\centering \cellcolor[HTML]{E3E2E2}\textcolor{blue}{$\checkmark$}}\\
		
		\hline
		
		\multicolumn{1}{c}{\centering 4} &
		\multicolumn{1}{c}{\centering \makecell{Huang'14\cite{huang14}, Li'14\cite{guo14}, Ali'15\cite{ali15}, \\Narayana'16\cite{narayana16}, Taherin'18~\cite{taherin15}, Awan'16\cite{Awan2016}}} &
		\multicolumn{1}{c}{\centering \textcolor{red}{$\times$}} &
		\multicolumn{1}{c}{\centering \textcolor{blue}{$\checkmark$}} &
		\multicolumn{1}{c}{\centering \textcolor{red}{$\times$}} &
		\multicolumn{1}{c}{\centering \textcolor{red}{$\times$}} &
		\multicolumn{1}{c}{\centering \textcolor{blue}{$\checkmark$}} &
		\multicolumn{1}{c}{\centering \textcolor{red}{$\times$}} &
		\multicolumn{1}{c}{\centering \textcolor{red}{$\times$}}\\
		
		\hline

		\multicolumn{1}{c}{\centering \cellcolor[HTML]{E3E2E2}5} &
		\multicolumn{1}{c}{\centering \cellcolor[HTML]{E3E2E2} Lee'14\cite{lee14}, Munawar'14~\cite{munawar14}} &
		\multicolumn{1}{c}{\centering \cellcolor[HTML]{E3E2E2}\textcolor{blue}{$\checkmark$}} &
		\multicolumn{1}{c}{\centering \cellcolor[HTML]{E3E2E2}\textcolor{red}{$\times$}} &
		\multicolumn{1}{c}{\centering \cellcolor[HTML]{E3E2E2}\textcolor{red}{$\times$}} &
		\multicolumn{1}{c}{\centering \cellcolor[HTML]{E3E2E2}\textcolor{red}{$\times$}} &
		\multicolumn{1}{c}{\centering \cellcolor[HTML]{E3E2E2}\textcolor{red}{$\times$}} &
		\multicolumn{1}{c}{\centering \cellcolor[HTML]{E3E2E2}\textcolor{red}{$\times$}} &
		\multicolumn{1}{c}{\centering \cellcolor[HTML]{E3E2E2}-}\\
		
		\hline
		\multicolumn{1}{c}{\centering \cellcolor[HTML]{E3E2E2}6} &
		\multicolumn{1}{c}{\centering \cellcolor[HTML]{E3E2E2} Lee'10~\cite{lee10}} &
		\multicolumn{1}{c}{\centering \cellcolor[HTML]{E3E2E2}\textcolor{blue}{$\checkmark$}} &
		\multicolumn{1}{c}{\centering \cellcolor[HTML]{E3E2E2}\textcolor{red}{$\times$}} &
		\multicolumn{1}{c}{\centering \cellcolor[HTML]{E3E2E2}\textcolor{red}{$\times$}} &
		\multicolumn{1}{c}{\centering \cellcolor[HTML]{E3E2E2}\textcolor{red}{$\times$}} &
		\multicolumn{1}{c}{\centering \cellcolor[HTML]{E3E2E2}\textcolor{red}{$\times$}} &
		\multicolumn{1}{c}{\centering \cellcolor[HTML]{E3E2E2}\textcolor{blue}{$\checkmark$}} &
		\multicolumn{1}{c}{\centering \cellcolor[HTML]{E3E2E2}-}\\
		
		\hline
		\multicolumn{1}{c}{\centering \cellcolor[HTML]{E3E2E2}7} &
		\multicolumn{1}{c}{\centering \cellcolor[HTML]{E3E2E2} Ansari'19~\cite{Ansari2018}} &
		\multicolumn{1}{c}{\centering \cellcolor[HTML]{E3E2E2}\textcolor{blue}{$\checkmark$}} &
		\multicolumn{1}{c}{\centering \cellcolor[HTML]{E3E2E2}\textcolor{blue}{$\checkmark$}} &
		\multicolumn{1}{c}{\centering \cellcolor[HTML]{E3E2E2}\textcolor{red}{$\times$}} &
		\multicolumn{1}{c}{\centering \cellcolor[HTML]{E3E2E2}\textcolor{red}{$\times$}} &
		\multicolumn{1}{c}{\centering \cellcolor[HTML]{E3E2E2}\textcolor{red}{$\times$}} &
		\multicolumn{1}{c}{\centering \cellcolor[HTML]{E3E2E2}\textcolor{blue}{$\checkmark$}} &
		\multicolumn{1}{c}{\centering \cellcolor[HTML]{E3E2E2}-}\\
		
		\hline
        \multicolumn{1}{c}{\centering 8} &
		\multicolumn{1}{c}{\centering Li'19~\cite{Li2019}} &
		\multicolumn{1}{c}{\centering \textcolor{red}{$\times$}} &
		\multicolumn{1}{c}{\centering \textcolor{blue}{$\checkmark$}} &
		\multicolumn{1}{c}{\centering \textcolor{blue}{$\checkmark$}} &
		\multicolumn{1}{c}{\centering \textcolor{red}{$\times$}} &
		\multicolumn{1}{c}{\centering \textcolor{blue}{$\checkmark$}} &
		\multicolumn{1}{c}{\centering \textcolor{red}{$\times$}} &
		\multicolumn{1}{c}{\centering \textcolor{red}{$\times$}}\\
		
		\hline
		\multicolumn{1}{c}{\centering \cellcolor[HTML]{E3E2E2}9} &
		\multicolumn{1}{c}{\centering \cellcolor[HTML]{E3E2E2} Ranjbar'19~\cite{Ranjbar19}, Ranjbar'20~\cite{Ranjbar2020}} &
		\multicolumn{1}{c}{\centering \cellcolor[HTML]{E3E2E2}\textcolor{red}{$\times$}} &
		\multicolumn{1}{c}{\centering \cellcolor[HTML]{E3E2E2}\textcolor{blue}{$\checkmark$}} &
		\multicolumn{1}{c}{\centering \cellcolor[HTML]{E3E2E2}\textcolor{blue}{$\checkmark$}} &
		\multicolumn{1}{c}{\centering \cellcolor[HTML]{E3E2E2}\textcolor{red}{$\times$}} &
		\multicolumn{1}{c}{\centering \cellcolor[HTML]{E3E2E2}\textcolor{blue}{$\checkmark$}} &
		\multicolumn{1}{c}{\centering \cellcolor[HTML]{E3E2E2}\textcolor{blue}{$\checkmark$}} &
		\multicolumn{1}{c}{\centering \cellcolor[HTML]{E3E2E2}\textcolor{blue}{$\checkmark$}}\\

		\hline
        \multicolumn{1}{c}{\centering 10} &
        \multicolumn{1}{c}{\centering Our Work} &
		\multicolumn{1}{c}{\centering \textcolor{blue}{$\checkmark$}} &
		\multicolumn{1}{c}{\centering \textcolor{red}{$\times$}} &
		\multicolumn{1}{c}{\centering \textcolor{blue}{$\checkmark$}} &
		\multicolumn{1}{c}{\centering \textcolor{blue}{$\checkmark$}} &
		\multicolumn{1}{c}{\centering \textcolor{blue}{$\checkmark$}} &
		\multicolumn{1}{c}{\centering \textcolor{blue}{$\checkmark$}} &
		\multicolumn{1}{c}{\centering \textcolor{blue}{$\checkmark$}}\\
		\bottomrule
	\end{tabular}
	}
\end{table*}

MC systems are the subject of recent research due to the emergence of the Cyber Physical System (CPS). Table~\ref{Table: related} summarizes the recent studies with different target optimization objectives. Since our focus is on power management and fault-tolerance for dependent MC tasks, we only consider the works presented for MC or non-MC systems with a similar scope. 
There are some algorithms presented for independent tasks such as 
Earliest Deadline First with Virtual Deadline (EDF-VD) used in~\cite{Ranjbar2020Fantom}, or using different scheduling policies for different criticality levels~\cite{anderson09}. Hence, these algorithms are presented for independent periodic tasks and cannot be applied to the tasks with precedence constraints.
Besides, as can be seen, some papers, such as~\cite{baruah16,socci15,Socci19,Medina17,Li16,pathan18,Ranjbar19}, have considered periodic MC tasks with data dependency but none of them have considered fault occurrence possibilities and power management. Rows 1 and 2 show that the papers in this area have focused on the feasibility of schedules and meeting the timing constraints without considering power consumption or fault-tolerance. 
From the perspective of guaranteeing LC tasks' minimum service level, most existing MC scheduling algorithms (row 1) discard or degrade LC tasks when the system switches to the HI mode. It causes serious service interruption for LC tasks. Therefore, in addition to power management and fault-tolerance in MC systems, improving the QoS of LC tasks would be significant. 
In addition, the used MC task model in~\cite{Li16} is different from the popular dependent MC task model by defining a criticality level for each task graph, not for each task in a graph. On the other hand, there are a few works (row~3) that proposed a method to schedule dependent MC tasks in multi-core systems~\cite{Medina18,Medina181}. These papers have considered fault occurrence possibilities, while they have not considered power or hotspot management. In terms of fault tolerance, researchers in~\cite{das2012} have proposed a design-time task re-mapping approach to tolerate faults, however, these techniques have not considered power management.


Research works have recently been presented on energy and power management in multi-core MC systems by considering periodic independent tasks (row 4). Most of these papers use DVFS technique to manage power consumption in the LO mode, and when the system enters into the HI mode, all HC tasks are executed with the high frequency, and also all of the LC tasks are dropped. Indeed, they interrupt the minimum service level of LC tasks in the HI mode. In addition, due to the high frequency in the HI mode, the system's peak power consumption may violate the TDP constraint. None of these papers that studied MC systems managed instantaneous power, especially when the system switches to the HI mode. Besides, researchers in~\cite{das2014,dasacm} have proposed a design-time task remapping approach to minimize the average power and tolerate faults; however, in addition to not managing the peak power in this method, it is not suitable for MC systems. 
On the other hand, some papers focused on peak power management of multi-core real-time systems (5-7). These works have proposed hard real-time tasks with one criticality level which is not practical for MC tasks. Although the authors in~\cite{Ansari2018} manage the peak power for the dependent task model, they use DVFS to manage the peak power consumption. Hence, it is not suitable for MC tasks, especially in the HI mode.

From the MC systems' thermal management perspective, researchers in~\cite{Li2019} have considered thermal management for independent MC tasks in single-core processors (row 8). Besides, researchers in~\cite{Ranjbar19,Ranjbar2020} reduce the peak power and manage the temperature in MC systems with the dependent task model (row 9). these papers use the DVFS technique, which is not acceptable, especially in the HI mode, and also, they have not considered fault occurrence. There is also no guarantee to manage peak power under TDP in their methods.

In this work, we study peak power and thermal management for MC multi-core systems by considering fault-tolerance techniques, which is not considered in existing MC works.


\section{System Models and Problem Definition}
\label{sec:models}

\subsection{Mixed-Criticality Task Model}
\label{subsec:mixedCriticalityTaskModel}

The MC system is responsible for executing application \textit{A} consisting of multiple dependent periodic tasks. Analogous to~\cite{baruah16,Medina18,Medina181,Ranjbar2020}, the application is modeled as a~directed acyclic graph $\textit{G}_{A}$ ($\textit{V}_{A}, \textit{E}_{A}$) in which each task in the~application is either a HC or a LC task. 
Each node $\textit{T}_{i}\in\textit{V}_{A}$~represents a task and an edge $\textit{e}_{ij}\in\textit{E}_{A}$ from $\textit{T}_{i}$ to $\textit{T}_{j}$ indicates a dependency between $\textit{T}_{i}$ and $\textit{T}_{j}$. A task is released and ready to be executed if all its predecessor tasks have finished their execution.
We assume preemptive execution for tasks, which means the tasks are interrupted during their execution on a core that mapped on it. 
Each task $\textit{T}_{i}$ is defined as: 
\begin{align}
	\textit{T}_{i} &= (\zeta_{i}, \textit{C}_{i}^{LO}, \textit{C}_{i}^{HI}, \textit{d}_{i}, \textit{P}_{i})
\end{align}	


Parameter $\zeta_{i}$ denotes levels of criticality for each task (HC or LC). 
Each task has a deadline ($\textit{d}_{i}$), low WCET ($\textit{C}_{i}^{LO}$), and high WCET ($\textit{C}_{i}^{HI}$). 
For each LC task, $\textit{C}_{i}^{LO}$ is equal to $\textit{C}_{i}^{HI}$, and for each HC task, $\textit{C}_{i}^{LO}$ is less than $\textit{C}_{i}^{HI}$.
The communication time between tasks is considered as a part of the predecessor task's execution time. 
All tasks in application \textit{A} have an identical period ($\textit{P}_{i}$), which is equal to the period of the application ($\textit{P}_{A}$)~\cite{salehi16, Ranjbar19}. 
Also, we assumed the deadline and period of the task graph are equal for an application~($\textit{d}_{A} = \textit{P}_{A}$). A deadline $d_i$ is determined for each task in order that all its successors can be scheduled before their deadlines. Hence, the deadlines of tasks that have no successors are equal to the task graph's deadline. 

\behnaz{\textbf{MC system's operational model:} MC systems first start the operation in the LO mode; if the execution time of at least one HC task exceeds its low WCET~($\textit{C}_{i}^{LO}$), the system switches to the HI mode. It stays in this mode until there is no ready HC task in each core's queue~\cite{baruah16,Medina18,Medina181,Ranjbar19}.} 
In the LO mode, the mapping and scheduling algorithms consider the low WCET of tasks while in the HI mode, the algorithm schedules tasks by their high WCETs.

\begin{figure}[t]
	\includegraphics[width=\linewidth]{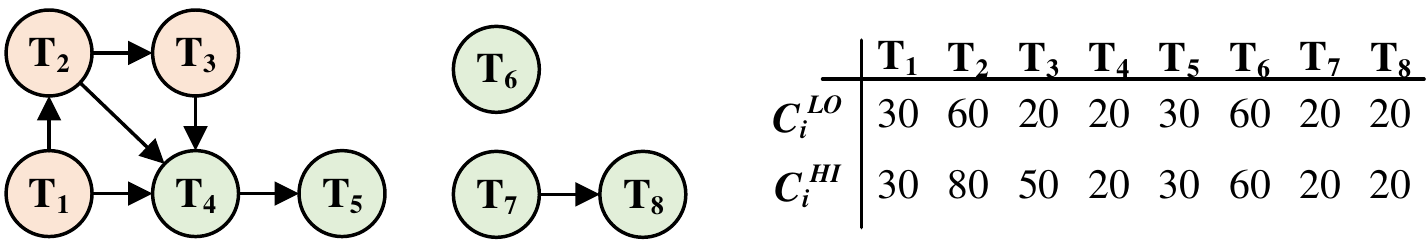}
	\caption{An example of real-life application task graph (Unmanned Air Vehicle)}
	\label{fig: reallife application}
\end{figure}

The task graph model is popular for image processing in automotive systems and pedestrian detection~\cite{izosimov10}. System designers assign the criticality level of tasks based on their functionalities. 
However, similar to previous studies in the literature~\cite{baruah16,Medina18}, if an LC task is a predecessor of an HC task, then it is considered as an HC task.
Fig.~\ref{fig: reallife application} shows the task graph of Unmanned Air Vehicle (UAV), which is a real-life MC application task graph~\cite{Medina18}.
This application is composed of eight tasks.
Tasks $\textit{T}_{1}$ to $\textit{T}_{3}$ are HC tasks that are responsible for the avoidance, navigation, and stability of the system.
Failure in the execution of these tasks before their deadline may lead to a system failure, and irreparable damage to the system.
The roles of LC tasks~($\textit{T}_{4}$ to $\textit{T}_{8}$), are recording sensors data, GPS coordination, and video transmissions~\cite{Medina18}.
The system should execute these tasks to improve its QoS; however, the system can skip executing them in harsh situations. Note that, the QoS is defined as the percentage of executed LC tasks in the HI mode to all LC tasks~\cite{li18,Ranjbar2020Fantom} ($QoS= n_L^{succ}/n_L$, where $n_L$ is the number of all LC tasks in a graph and $n_L^{succ}$ is the number of executed LC tasks that, when the system switches to the HI mode and $Finish-time_i\leq d_i \And \zeta_i= LC$). 
\behnaz{Dropping some LC tasks in the HI mode can be used for real-time applications characterized by hard and firm deadlines. 
The tasks with a hard deadline can be HC tasks, and with firm deadlines 
can be LC tasks. The multimedia tasks are an example of firm deadlines, where skipping a video frame once in a while is better than processing it with a long delay or not processing it completely~\cite{buttazzo11}. }

\subsection{Fault-Tolerance Model}
\label{subsec:faultTolerantModel}
Transient faults are the most common faults in embedded systems~\cite{guo14,Ranjbar2020Fantom,sahoo2021reliability}. \behnaz{To tolerate transient faults, fault detection and correction mechanisms need to be applied. The detail of these mechanisms is discussed in Appendix A.}

\behnaz{For embedded safety-critical real-time systems, low-cost, low power, and high accuracy checker should be employed in each core. 
To check whether a fault occurs during the execution of a task, analogous to~\cite{salehi16,Ansari2018,Ranjbar2020Fantom}, an error detection mechanism, 
is conducted to check the correctness of the task's output at the end of the task's execution. ARGUS~\cite{Meixner2007} is one of the significant checker tools to detect errors, that has all the features and has been used in many recent works~\cite{Cerrolaza2020}. It can be applied to any embedded systems with less than 11\% chip area overhead and also check control flow, dataflow, computation, and memory access separately, at run-time. 
Here, the error detection time overhead is considered in the WCET of tasks. Besides, the task re-execution technique is one of the most popular ways to correct transient faults in embedded systems~\cite{Ranjbar2020Fantom, koren07}, which we employed in this article. }
We assume that up to~\textit{k} transient faults may occur in one period of the application~\cite{Izosimov08,izosimov10,Salehi2016}. If the system detects a faulty task, it spends some time~($\mu$) to discard the results of the faulty task before re-executing the task.

\subsection{Power Model}
\label{PowModel}
The total power consumption of a core in an embedded system consists of three major components, Dynamic power~($P_{d}$), static power~($P_{s}$), and frequency-independent power~($P_{ind}$)~\cite{taherin15,Ranjbar19}. 
\answersnd{We consider a pessimistic approach to obtain the dynamic power consumption of HC tasks as well as LC tasks.}
$P_{d}$ is a frequency and workload dependent power which is consumed when a task is run. 
$P_{s}$ is consumed in a core even when the core is in idle mode and no task is run. 
The frequency-independent power is the power used for memory and I/O operations~\cite{taherin15,guo14}. 
The total power is:
\begin{equation}
\label{eq:Power} 
\begin{split}
P(V_i,f_i) &= P_{d} + P_{s} + P_{ind} = \alpha C_{eff}V_i^2f_i + I_{sub}V_{i} + P_{ind}
\end{split}
\end{equation}
where $\alpha$, $I_{sub}$, $C_{eff}$, $f_i$, and $V_i$ are transition rate, sub-threshold leakage current, effective capacitance, core $i$ processing frequency, and core $i$ supply voltage, respectively.

\subsection{Problem Definition}
\label{sec:problemFormulation}


\textit{Deadline Constraint:} Each HC task $\tau_i$, must finish its execution ($ft_i$) correctly before its deadline ($d_i$) in both LO and HI modes. In addition, all LC tasks should finish their execution before their deadlines in the LO mode.
\begin{align}
\forall \tau_i,& ~\zeta_i= HC: ft_i \leq d_i \nonumber \\
\forall \tau_i,& ~\zeta_i= LC \AND Cr_L= LO: ft_i \leq d_i 
\end{align}


\textit{Task Dependability Constraint:} Due to the precedence correlations between tasks, the start time of task $\tau_i$ ($st_i$) must be \behnaz{greater} than the finish time of all its predecessor tasks ($Pred (\tau_i)$).
\begin{align}
\forall \tau_i, \forall j\in Pred (\tau_i) \implies st_i \geq ft_j
\end{align}

\textit{Mapping Constraint:} A task ($\tau_i$) can only be executed on a single core in each time slot. If $X_{ij}$ denotes the mapping of task $\tau_i$ on core $j$, then:
\begin{align}
 \forall \tau_i, \sum_{j \in Cores}{X_{ij}} = 1  
\end{align}

\begin{table*}[ht]
	\centering
	\caption{All possible scenarios of executing the task graph presented in \figureRef{fig:a_taskGraph} on a single core chip}
	\begin{threeparttable}
		\begin{tabular}{ccccccccccccccc}
			\toprule
			\multirow{1}{1cm}{\centering \textit} &
			\multirow{1}{0.6cm}{\centering $S_1$ }&
			\multirow{1}{0.6cm}{\centering $S_2$ }&
			\multirow{1}{0.6cm}{\centering $S_3$ }&
			\multirow{1}{0.6cm}{\centering $S_4$ }&
			\multirow{1}{0.6cm}{\centering $S_5$ }&
			\multirow{1}{0.6cm}{\centering $S_6$ }&
			\multirow{1}{0.6cm}{\centering $S_7$ }&
			\multirow{1}{0.6cm}{\centering $S_8$ }&
			\multirow{1}{0.6cm}{\centering $S_9$ }&
			\multirow{1}{0.6cm}{\centering $S_{10}$ }&
			\multirow{1}{0.6cm}{\centering $S_{11}$ }&
			\multirow{1}{0.6cm}{\centering $S_{12}$ }&
			\multirow{1}{0.6cm}{\centering $S_{13}$ }&
			\multirow{1}{0.6cm}{\centering $S_{14}$ }\\
			\midrule
			\multicolumn{1}{c}{Overrun}	&\multicolumn{1}{c}{-}	&\multicolumn{1}{c}{-}	&\multicolumn{1}{c}{-}	&\multicolumn{1}{c}{-}	&\multicolumn{1}{c}{$T_1$}	&\multicolumn{1}{c}{$T_2$} &\multicolumn{1}{c}{$T_1$\tnote{*}~\raisebox{.5pt}{\textcircled{\raisebox{-.9pt} {1}}}}	&\multicolumn{1}{c}{$T_1$\raisebox{.5pt}{\textcircled{\raisebox{-.9pt} {1}}}} 	&\multicolumn{1}{c}{$T_1$\raisebox{.5pt}{\textcircled{\raisebox{-.9pt} {1}}}}	&\multicolumn{1}{c}{$T_2$\raisebox{.5pt}{\textcircled{\raisebox{-.9pt} {1}}}}  &\multicolumn{1}{c}{$T_2$\raisebox{.5pt}{\textcircled{\raisebox{-.9pt} {1}}}}	&\multicolumn{1}{c}{$T_1$\raisebox{.5pt}{\textcircled{\raisebox{-.9pt} {2}}}}	&\multicolumn{1}{c}{$T_2$\raisebox{.5pt}{\textcircled{\raisebox{-.9pt} {2}}}} 	&\multicolumn{1}{c}{$T_2$\raisebox{.5pt}{\textcircled{\raisebox{-.9pt} {2}}}}  \\
			\multicolumn{1}{c}{Fault}					& \multicolumn{1}{c}{-}	& \multicolumn{1}{c}{$T_1$}	&	\multicolumn{1}{c}{$T_2$}	& \multicolumn{1}{c}{$T_3$}	& \multicolumn{1}{c}{-}		& \multicolumn{1}{c}{-} 	& \multicolumn{1}{c}{$T_1$\raisebox{.5pt}{\textcircled{\raisebox{-.9pt} {2}}}}	& \multicolumn{1}{c}{$T_2$\raisebox{.5pt}{\textcircled{\raisebox{-.9pt} {2}}}} 	& \multicolumn{1}{c}{$T_3$\raisebox{.5pt}{\textcircled{\raisebox{-.9pt} {2}}}}	& \multicolumn{1}{c}{$T_2$\raisebox{.5pt}{\textcircled{\raisebox{-.9pt} {2}}}}  & \multicolumn{1}{c}{$T_3$\raisebox{.5pt}{\textcircled{\raisebox{-.9pt} {2}}}}	& \multicolumn{1}{c}{$T_1$\raisebox{.5pt}{\textcircled{\raisebox{-.9pt} {1}}}}	& \multicolumn{1}{c}{$T_1$\raisebox{.5pt}{\textcircled{\raisebox{-.9pt} {1}}}} 	& \multicolumn{1}{c}{$T_2$\raisebox{.5pt}{\textcircled{\raisebox{-.9pt} {1}}}} 	\\
			\multicolumn{1}{c}{Finish time}	&\multicolumn{1}{c}{9}	& \multicolumn{1}{c}{14}	&	\multicolumn{1}{c}{13}		& \multicolumn{1}{c}{12}	&\multicolumn{1}{c}{13}	& \multicolumn{1}{c}{11} & \multicolumn{1}{c}{\textbf{20}\tnote{***}}	 													& \multicolumn{1}{c}{\textbf{19}\tnote{***}} 														& \multicolumn{1}{c}{16}														& \multicolumn{1}{c}{17}  							& \multicolumn{1}{c}{12}	& \multicolumn{1}{c}{18}	& \multicolumn{1}{c}{16} 	& \multicolumn{1}{c}{15}  \\
			\multicolumn{1}{c}{Exe time (drop $T_3$)}	& \multicolumn{1}{c}{7}	 	& \multicolumn{1}{c}{12}	&	\multicolumn{1}{c}{11}		& \multicolumn{1}{c}{10\tnote{**}}	& \multicolumn{1}{c}{11}	& \multicolumn{1}{c}{9} & \multicolumn{1}{c}{18}	 & \multicolumn{1}{c}{17} 			& \multicolumn{1}{c}{14\tnote{**}}		& \multicolumn{1}{c}{15}  							& \multicolumn{1}{c}{10\tnote{**}}	& \multicolumn{1}{c}{16}	& \multicolumn{1}{c}{14} 	& \multicolumn{1}{c}{13}  \\
			\bottomrule
		\end{tabular}%
		\begin{tablenotes}
			\item[*] {\footnotesize The circles shows the order of the occurrence of fault and task overrun.}
			\item[**] {\footnotesize In these cases, the system executes $T_3$ and detects a fault occurs, but does not re-execute the task (drops the task).}
			\item[***] {\footnotesize In these cases, the system must drop task $T_3$ to execute all HC tasks before their deadlines.}
		\end{tablenotes}
	\end{threeparttable}
	\label{tab:Motivational}%
\end{table*}%

\textit{Power Constraint:} The chip's overall power consumption must not violate the chip's TDP~($TDP_{chip}$) in any time slot.
\begin{align}
 & \forall t \in timeslots: \sum_{j \in Cores}{Pow_{jt}} \leq TDP_{chip},  
\end{align}
where $Pow_{jt}$ represents the power consumption of core $j$ in time slot $t$.

When the system switches to the HI mode, the system drops some LC tasks to meet the timing constraints, which degrades the QoS of the system:
\begin{align}
 &Cr_L= HI: \quad QoS = n_L^{succ}/n_L 
\end{align}

The problem is how to map and schedule dependent MC tasks of application A on the system's cores to satisfy the aforementioned constraints (timing and peak power) and QoS of the system.
In this paper, we propose a heuristic method to solve this NP-hard problem \cite{munawar14}.


\section{PROPOSED METHOD}
\label{sec:proposalMethod}

In this section, at first, a motivational example is presented in Section~\ref{subsec:motivationalExample} for a better understanding of the problem and the proposed solution. 
Then, the proposed method is explained in detail in Section~\ref{subsec:Method} and~\ref{subsec:treeConstructionStrategies}.

\subsection{Motivational Example}
\label{subsec:motivationalExample}

\begin{figure}[!t]
	\centering
	\includegraphics[width=1\columnwidth]{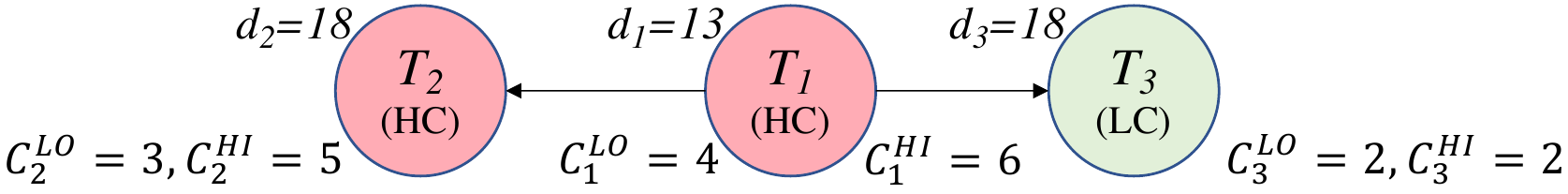}
	\caption{Task graph of an application with three tasks used in the example}
	\label{fig:a_taskGraph}
\end{figure}

\figureRef{fig:a_taskGraph} shows an application with three tasks, where tasks ${T}_{1}$ and ${T}_{2}$ have HC, and task ${T}_{3}$ has LC.
Deadline, low WCET, and high WCET of each task are presented in the figure, and the system takes 1ms to discard the output of a faulty task ($\mu=1$). Hence, the period of all tasks are the same and is equal to \textit{18}ms. 
For the sake of simplicity, we considered that the application runs on a single-core processor, and up to one fault may occur during the execution of the application ($k=1$). 
So, the system cannot execute multiple tasks simultaneously on different cores to violate TDP (we will discuss TDP challenge later). 
The scheduling algorithm in the LO mode, executes tasks $T_1$, $T_2$, and $T_3$, respectively. 
The schedulability test~\cite{baruah16} shows that in the LO mode, all three tasks can be executed even in the case of fault occurrence. In other words, the total CPU utilization for application $A$ is less than one $(U_{A} \leq 1)$. 
If we consider the HI mode, if a LC task $T_3$ is executed in addition to the two HC tasks in the case of fault occurrence, the system becomes overloaded~($U_A^{HI}$>1) and the three tasks cannot be scheduled. However, if we drop some LC tasks in the HI mode ($T_3$ in this example) to guarantee the correct execution of HC tasks, then the computation demand requested by tasks is less than one and can be scheduled before their deadline. As a result, the utilization of this example for both LO and HI modes with the probability of one fault occurrence is computed as follows, in which, just LC tasks are considered to be executed in the HI mode.
\begin{align}
\label{eq:3}
\begin{split}
U_A &= MAX(U_A^{LO}, U_A^{HI}) \leq 1 \\
U_A^{LO} &= (\sum_{i \in \lbrace 1,2,3\rbrace} \frac{C_i^{LO}}{P_A})
+\frac{k(\max_{i \in \lbrace 1,2,3\rbrace}(C_i^{LO}) +\mu)}{P_A}\\
&=\frac{4}{18}+\frac{3}{18}+\frac{2}{18} +(\frac{4+1}{18})=\frac{14}{18}<1 \\
U_A^{HI} &= (\sum_{i \in \lbrace 1,2\rbrace} \frac{C_i^{HI}}{P_A})+\frac{k(\max_{i \in \lbrace 1,2\rbrace}(C_i^{HI}) +\mu)}{P_A}\\
&=\frac{6}{18}+\frac{5}{18}+(\frac{6+1}{18})=\frac{18}{18} \leq 1
\end{split}
\end{align}

However, for this example, fourteen different scenarios could happen during the execution of the application because the time of the fault and task overruns are unknown.
\tableRef{tab:Motivational} shows all these scenarios, and the execution time of the system whether it drops LC task or not.
In ten scenarios ($S_5$ to $S_{14}$), an HC task overruns, and it shows the system is in the HI mode.
However, as shown in \tableRef{tab:Motivational}, only in two scenarios, the system fails to execute all tasks (HC and LC tasks) before the deadline ($S_7$ and $S_8$). The reason is that, the system has switched to the HI mode in these scenarios and also a fault has occurred. it causes the system to be overloaded and the computation demand for executing all tasks becomes more than one~($U_A>1$). Therefore, the LC task $T_3$ would be dropped. Although there are some scenarios such as $S_{12}$ to $S_{14}$, that the system is in the HI mode, we schedule the LC task in this mode to improve the QoS. As shown in \tableRef{tab:Motivational}, the start time of $T_3$ in $S_{12}$ ($S_{14}$) is 16 (13), and since the WCET of the LC task is 2, then it can be executed before the application deadline ($d_A=18$).
This example clearly shows that all situations should be considered in an MC system. Therefore, the system is analyzed in detail at design-time, and then, the proper schedule is exploited in the online phase to minimize the drop ratio of LC tasks and enhance the QoS.

\begin{figure}
	\centering
	\includegraphics[width=0.73\columnwidth]{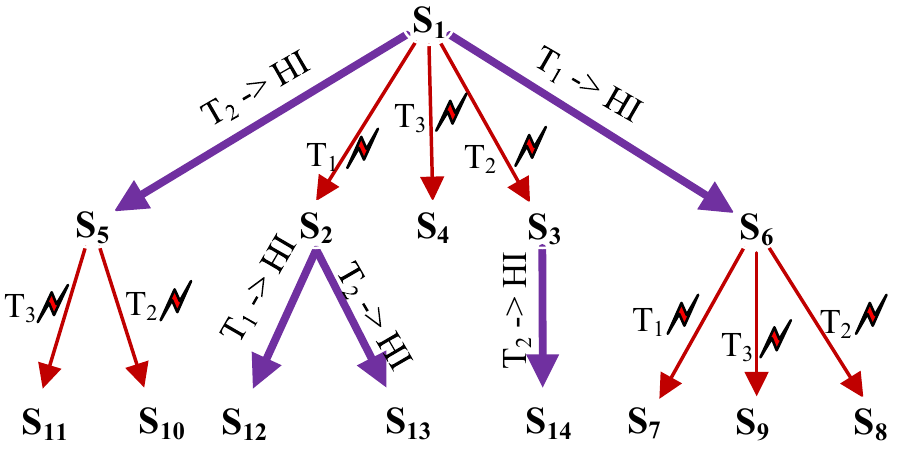}
	\caption{The tree constructed by our method for the task graph of \figureRef{fig:a_taskGraph}}
	\label{fig:b_scheduleTree}
\end{figure}

\figureRef{fig:b_scheduleTree} shows the tree for the application task graph presented in the motivational example, which is constructed in the offline phase of our proposed approach.
At run-time, the system starts each period with $S_1$ (the scheduling in the root of the tree), which corresponds to the scenario where no fault occurs and no HC tasks overruns. 
If an HC task overruns (for instance task $T_1$), The system searches through the children of the current node ($S_1$), finds the appropriate task mapping and scheduling, and continues the execution based on the new schedule ($S_6$ in this case). 
After that, if error detection unit detects a fault at the end of a task execution (for instance task $T_2$), the system searches through the children of the current node ($S_6$), finds the appropriate scenario, and continues the execution based on the new task mapping and scheduling ($S_{14}$ in this case).

It is important to mention that each schedule has a different start time, system mode, the expected number of faults, and task set.
Furthermore, scheduling of child nodes must be compatible with the scheduling of their parent; so, the system can change the schedule of tasks without any conflicts. 
For instance, assume that in $S_1$, the task execution order is $T_1$, $T_2$, and $T_3$. 
When the system employed $S_4$, it implies that $T_1$ and $T_2$ are completed successfully, the system is in the LO mode, and a fault is detected at the end of the first execution of $T_3$. 
So, the $S_4$ should schedule tasks based on this information.


\subsection{Design Methodology}
\label{subsec:Method}

\begin{figure}
	\centering
	\includegraphics[width=1\columnwidth]{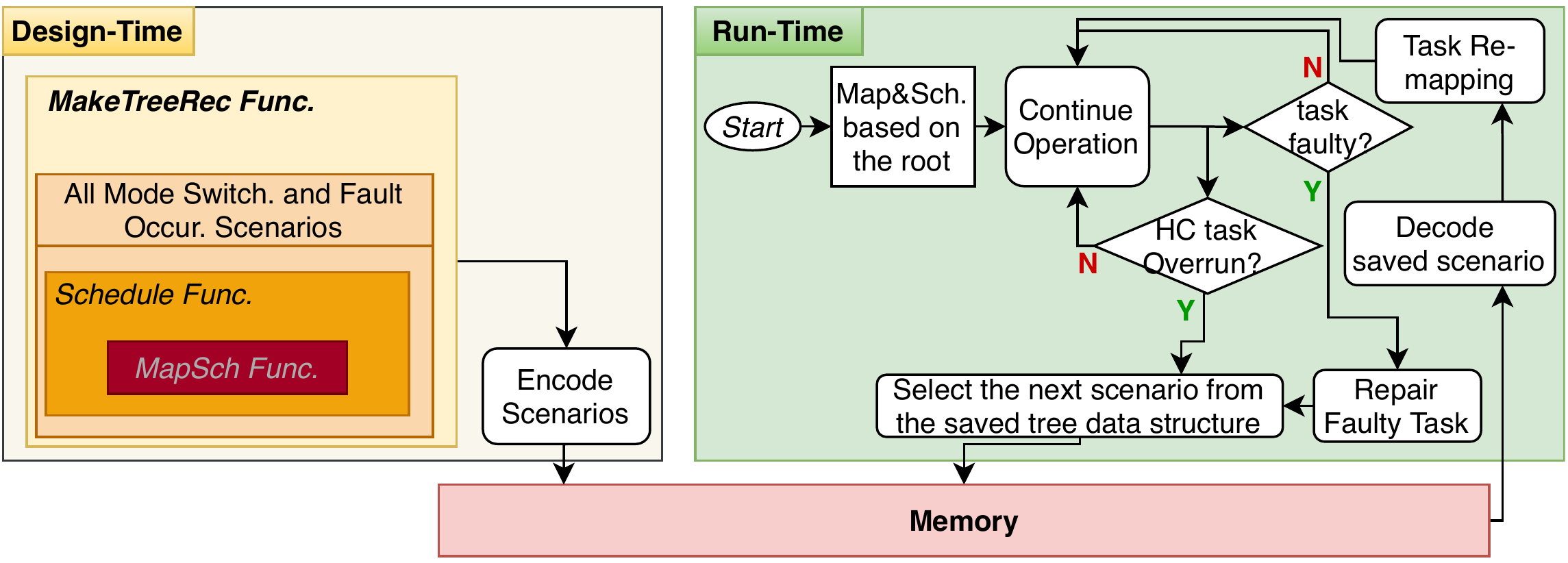}
	\caption{Design methodology}
	\label{fig:method}
\end{figure}

The fault-tolerance and peak power-aware task mapping and scheduling method consists of two phases, design-time, and run-time. In this paper, we focus on system design at design-time. For the sake of completeness, we provide a brief overview of how to use the schedule tree in the run-time phase, which is generated at design-time. Fig.~\ref{fig:method} shows an overview of the design methodology. In the design-time phase, there are three functions which are used to generate the schedule tree, \textit{MakeTreeRec}, \textit{Schedule}, and \textit{MapSch}. Section~\ref{subsec:treeConstructionStrategies} provides details of generating the tree, these functions, and how we manage the peak power consumption. All scenarios are stored in memory to be used in the run-time phase. At run-time, an application follows the presented task mapping and scheduling in the root of the tree. In the case of fault occurrence or mode switching, the appropriate task mapping and scheduling for the remaining un-executed tasks are fetched from memory. After fetching, mapped tasks based on the previous scenario are re-mapped based on the new scenario, and the system continues its operation. In the following subsection, we explain the design-time phase of our proposed method.

\subsection{Tree Generation and Fault-Tolerant Scheduling \& Mapping}
\label{subsec:treeConstructionStrategies}

As we discussed, multiple scenarios might happen during the execution of an instance of the application, where, in most of these scenarios, the system can execute all or most of LC tasks without violating HC tasks' deadline.
To this end, the proposed approach of this paper considers a different mapping and scheduling for each scenario to handle HC tasks deadlines, faults, and peak power violations while minimizing the number of dropped LC tasks in the HI mode.
In the run-time phase, in general, the system is unaware of tasks that might overrun or a fault occurs; so, the system cannot select the proper schedule in advance.
Therefore, this paper employs a tree data structure in the offline phase to organize the mapping and scheduling of tasks for all scenarios, corresponding to~each HC task overrunning, and/or up to $k$ faults occurrence during each period. Now, we explain how the scheduling tree is generated.

\subsubsection{\textbf{Making Scheduling Tree}}

The main function of creating the tree ($\Phi$) is outlined in Algorithm~\ref{alg:schedulingTree}.
At first, we define a priority queue called \textit{TaskPQ} (line 3), which considers tasks' release time as the priority.
The release time of a task is the time when all its predecessor tasks have finished their execution (presented in Section \Ref{subsec:mixedCriticalityTaskModel}).
Then, the algorithm enqueues all tasks without any predecessor to the \textit{TaskPQ} with key equal to 0; because they are released at the beginning of the period (line 4).
Each node of the tree represents a particular scenario, and it has two attributes called \textit{sch} and \textit{childs}.
For instance, in the root scenario, the system is in the LO mode, and no fault occurs in the entire period.
\textit{sch} is the proper mapping and scheduling of tasks for that scenario, and \textit{childs} is a list of children nodes of the current node.

\begin{algorithm}[!t]
\footnotesize
		\caption{Creating the Tree}
		\label{alg:schedulingTree}
		\begin{threeparttable}
				\begin{algorithmic}[1]
					\renewcommand{\algorithmicrequire}{\textbf{Input:}}
					\renewcommand{\algorithmicensure}{\textbf{Output:}}
					
					\Require Task Graph ($G_{T}$), List of Cores ($c$), Number of Faults ($k$).
					\Ensure Scheduling Tree ($\Phi$).

					\Procedure{MakeTreeMain}{}
					\State $\Phi \gets$ Empty Tree
					\State \textit{TaskPQ} $\gets$ Empty Priority Queue
					\State Add all tasks without predecessor node to \textit{TaskPQ} with key $0$
					\State $\Phi[root].sch \gets MapSch (G_{T}, c, 0, \varnothing, TaskPQ)$ 
					\If {$\Phi[root].sch$=\textit{un-scheduled}}
					\State \Return {$\varnothing$}
					\EndIf
					\State $\Phi[root].childs \gets MakeTreeRec (G_{T}, c, $\algbrakline{4.5}$\Phi[root].sch, 0, k, LO)$
					\If {$\Phi[root].childs$=\textit{un-scheduled}}
					\State \Return {\textit{un-scheduled}}
					\EndIf
					\State \Return {$\Phi$}
					\EndProcedure
					
					\Function{MakeTreeRec}{Task Graph ($G_{T}$), List of Cores ($c$), Parent Schedule($Sch$), Time ($T$), Number of Faults ($k$), Mode of the System ($Mode$)}

					\State \textit{SchList} $\gets$ Empty list of nodes
					
					\If {\textit{Mode} = \textit{LO}} // \textit{HChild} nodes
					\State \textit{Tasks} $\gets$ List of unfinished HC tasks in time \textit{T}.
					\State $G_{H} \gets$ $G_{T}$ With High WCET.
										
					\ForEach {$\tau$ \textbf{in} \textit{Tasks}} 
					\State $T_{tmp} \gets$ Finish time of $\tau$ in \textit{sch} + $T_{sw}$.
					\State S $\gets$ New node
					\State S.sch = {\textit{Schedules}} ($G_{H}$, c, $T_{tmp}$, \textit{sch}, \textit{TaskPQ})
					\If {S.sch = un-scheduled}
					\State \textbf{return} $\varnothing$.
					\EndIf		
					\State $S.childs \gets MKTreeRec (G_{H}, c, S, T_{tmp}, k, HI)$
					\If {$S.childs = \varnothing$}
					\State \textbf{return} $\varnothing$.
					\EndIf	
					\State Add $S$ to $SchList$			 				
					
					\EndFor	
					\EndIf
					
					\If {\textit{k} $>$ 0}
					\State \textit{Tasks} $\gets$ List of Unfinished Tasks in \textit{T}.
					
					\ForEach {$\tau$ in \textit{Tasks}} 
					\State $T_{tmp} \gets$ Finish Time of $\tau_{i}$ in \textit{sch}.
					\State S.sch = {\textit{Schedules}}({$G_{T}$,~\textit{sch},~\textit{Time}})
					\If {S.sch = un-scheduled}
					\State \textbf{return} un-scheduled.
					\EndIf						
					
					\State $S.childs \gets MKTreeRec (G_{T},c,S, T_{tmp},k-1, Mode)$
					\If {$S.childs = \varnothing$}
					\State \textbf{return} $\varnothing$.
					\EndIf
					\State Add $S$ to $SchList$			 						
					
					\EndFor	
					\EndIf
					
					\State \Return{$SchList$}
					\EndFunction

			 	\end{algorithmic}
		\end{threeparttable}
\end{algorithm}

The algorithm calls \textit{MapSch} function (Algorithm~\ref{alg:mapsche}, which is discussed later in Section~\ref{subsubsec:MapandSch}) to schedule task for the root node (line~5).
The algorithm returns \textit{un-scheduled}, if \textit{MapSch} function cannot find any feasible schedule with no task dropping and violating the TDP constraint (lines 6-8).
Otherwise, the algorithm in line 9 continues to create the rest of the tree recursively by calling \textit{MakeTreeRec} function (which is presented in lines 15-50 of this algorithm).
If task scheduling is feasible in all possible scenarios, \textit{MakeTreeRec} function returns a list of child nodes, and the algorithm returns the tree ($\Phi$); otherwise, the algorithm returns \textit{un-scheduled} which means it could not find a feasible solution (lines 10-13).

The \textit{MakeTreeRec} function in Algorithm~\ref{alg:schedulingTree} recursively creates the tree.
Each node in the tree might have two types of child nodes.
The first type of child node (\textit{HChild}) has a scenario similar to their parents, except that one of the HC tasks overruns. 
Therefore, if the system is in the LO mode, any unfinished HC task might overrun and change the system's mode.
To this end, first, \textit{MakeTreeRec} function collects all unfinished HC tasks and creates an HC task graph ($G_H$) by changing the WCET of all tasks to high WCET (lines 18-19). 
Then, for each unfinished HC task, the function considers the scenario that the task overruns and schedules the task by calling \textit{Schedules} function, presented in Algorithm~\ref{alg:schedulingTree2}.
If the \textit{Schedules} function finds feasible scheduling, the algorithm recursively creates a tree for this node, where the system is in the HI mode, and up $k$ faults may occur on the remaining tasks by calling \textit{MakeTreeRec} function (lines 20-32). 
It is important to mention that switching to the HI mode has non-zero timing overhead ($T_{sw}$) in realistic systems~\cite{Chisholm17}, but it is insignificant in comparison with tasks' WCETs (line 21). 
The second type of child node (\textit{FChild}) has a scenario similar to their parents, except one fault occurs during the execution of one of the remaining tasks.
As we mentioned in \sectionRef{subsec:faultTolerantModel}, the system can tolerate up to $k$ faults in a period.
If less than \textit{k} faults occur in a node scenario, a faulty execution of all remaining tasks needs to be considered. 
Therefore, a child node is generated for the faulty execution of each remaining task, and also for each child node, the algorithm recursively constructs a tree by calling \textit{MKTreeRec} with \textit{k-1} faults (line 34-48).
Finally, if the algorithm finds a feasible solution for all scenarios, it returns the list of child nodes ($SchList$).

\begin{algorithm}[!t]
	\footnotesize
	\caption{Schedule Procedure}
	\label{alg:schedulingTree2}
	\begin{threeparttable}
		\begin{algorithmic}[1]
			\renewcommand{\algorithmicrequire}{\textbf{Input:}}
			\renewcommand{\algorithmicensure}{\textbf{Output:}}				

					\Require Task Graph ($G_{T}$), List of Cores ($Cores$), Time ($T$), Parent Schedule ($Sch_{par}$), Ready Task Priority Queue ($TaskPQ$).
					\Ensure Schedule ($Sch$)

					\Procedure{Schedules}{}			
					\State $T_{r} \gets$ List of Tasks from $G$ that all predecessors \algbrakline{1.15} has started executing before time \textit{T}.
					
					\State Add $T_{r}$ Tasks to a Priority Queue (\textit{TaskPQ}).
					
					\State $Sch \gets Sch_{par}[0-T]$
					\State $Sch \gets MapSch (G_1, Cores, T, Sch, TaskPQ)$
					\If {$Sch$ = un-scheduled}
					\State Find a LC Task with Largest execution time which has not started \algbrakline{0.8} in \textit{T} and remove it from $G$ then \textbf{goto} line 2.
					\If {Cannot find any LC task} 
					\State \textbf{return} un-scheduled.
					\EndIf
					\EndIf					
					\State \textbf{return} $Sch$
					\EndProcedure
			
		\end{algorithmic}
	\end{threeparttable}
\end{algorithm}

The \textit{Schedule} function, schedules tasks for each situation by calling \textit{MapSch} function (Algorithm~\ref{alg:mapsche}).
If \textit{MapSch} function fails to find a feasible solution to meet the deadlines of all tasks with respect to TDP constraint, \textit{Schedules} function drops the largest LC task (in terms of WCET) and calls \textit{MapSch} function again.
The \textit{Schedules} function repeats this procedure to find a feasible schedule.
If \textit{MapSch} function fails to find a feasible solution, and there is no more LC task to drop, \textit{Schedules} function returns \textit{un-scheduled} (lines 2-12 Algorithm~\ref{alg:schedulingTree2}).
We will discuss the \textit{MapSch} function in the next subsection.
As we mentioned in this section, occurring faults and a criticality mode change generate different scheduling scenarios that correspond to a set of alternative schedules.
These scenarios are stored in the memory of the system as a tree in the offline phase. 
At run-time, the system starts with the scheduling in the root node, which is for the scenario that no fault or overrun happens.
After that, if a fault occurs or an HC task overruns, the system finds the appropriate scenario in the child nodes of the current node and changes the scheduling of the system to improve the number of executed LC tasks.

\begin{algorithm}[!t]
\footnotesize
		\caption{Mapping and Scheduling Pseudo Code}
		\label{alg:mapsche}
		\begin{threeparttable}
			{ \small
				
				\begin{algorithmic}[1]
					\renewcommand{\algorithmicrequire}{\textbf{Input:}}
					\renewcommand{\algorithmicensure}{\textbf{Output:}}
					
					\Require Task Graph ($G_{T}$), List of Cores ($Cores$), Time ($T$), Scedule up to the Time ($Sch$), Ready Task Priority Queue ($TaskPQ$).
					
					\Ensure Complete Schedule ($Sch$)
					
					\Procedure{MapSch}{}
					    
					    \For {\textit{TS} = $T$ \textbf{to} \textit{PERIOD}}
					        \State $ReadyTasks \gets \varnothing$
					        
					        \State Extract minimum element from \textit{TaskPQ} and add it
					        to  \par
					        \hskip\algorithmicindent
					        \textit{ReadyTasks} while key of each element is equal to \par
					        \hskip\algorithmicindent \textit{TS} and \textit{TaskPQ} is not empty;
					        
					        \If{\textit{TaskPQ}.empty() = \textbf{true} \textbf{and} \textit{ReadyTasks} $= \varnothing$}
					        \State \Return $Sch$ \textit{//} Scheduling is done;
					        \EndIf

					        \If{\textit{ReadyTasks} $= \varnothing$}
					            \State \textbf{continue} \textit{//} No new task is ready in this \textit{TimeSlot}
					       \EndIf

					      \State \textit{SortedTasks} $\gets$ Sort (\textit{ReadyTasks, Desc}); 
                          \State \textit{SortedCores} $\gets$ Sort (\textit{Cores, Asc});

                                \For {\textit{task} \textbf{in} \textit{SortedTasks}}
                                    \For {\textit{core} \textbf{in} \textit{SortedCores}}
                               \State $\textit{Time}_\textit{tmp} \gets \textit{task}_\textit{wcet}$
                               
                               \State $count \gets 0$

                               \State $Sch_{tmp} \gets Sch$
                               
                               \State $SysPow_{tmp} \gets SysPow$

                                \While {$Time_{tmp}$ > $0$}
                                 
                                \If{$Sch_{tmp}(TS$+$count, core)$ is empty \& \algbrakline{1.2} $SysPow_{tmp}(TS$+$count)$ + $task_{pow}$ <= $TDP$}
                                
                                   \State $Sch_{tmp}(TS+count, core)$ = $task$
                                   
                                   \State $SysPow_{tmp}(TS+count)$ += $task_{pow}$
                                   
                                   \State $Time_{tmp}$ -= $1$ 
                                    
                                \EndIf
                                \State $count$ += $1$;
                                \EndWhile

                              \If{$TS$+$count \leq task_{deadline}$} 
                                \State $Sch \gets Sch_{tmp}$
                               
                               \State $SysPow \gets SysPow_{tmp}$

                                
                                \State \textit{SortedCores} $\gets$ Sort (\textit{Cores, Asc}); 
                         \State $task_{sch} \gets $ \textbf{true}
                                \State \textbf{break}
                             \EndIf 
                                
                                    \EndFor

                             \If{$task_{sch}$ == \textbf{false}}
  
                                           \State \Return {un-scheduled}

                                        \EndIf

                                \EndFor

				    \EndFor                        
					\EndProcedure
				\end{algorithmic}
			}
		\end{threeparttable}
\end{algorithm}

\subsubsection{\textbf{Mapping and Scheduling}}
\label{subsubsec:MapandSch}
In this section, we explain the proposed mapping and scheduling algorithm, which manages the peak power and hotspot distribution.
It should be mentioned that low power techniques, e.g., DVFS, cannot be easily used in the HI mode, especially when the system is in the overload situation due to the timing overhead. Therefore, we manage the peak power by finding the proper mapping and scheduling of tasks on free time slots of cores. This task mapping and scheduling are feasible if the system's power consumption never exceeds the TDP constraint, and all tasks finish their execution (even in the worst case) before their deadline.
So, \textit{MapSch} algorithm decides the time and core where each task should be executed.


Algorithm~\ref{alg:mapsche} outlines the pseudo-code of the \textit{MapSch} algorithm.
\behnaz{Tasks are mapped and scheduled up to time \textit{T} based on the current node schedule. In the case of fault occurrence or mode switches at time \textit{T}, this algorithm maps and schedules the rest of the tasks based on the new node schedule from time \textit{T} to the end of the application period (\textit{PERIOD}).}
The time is divided into a set of equal time slots (\textit{TS}), and the scheduler will put tasks into cores only at the beginning of each time slot.
In each time slot, at first, the algorithm sets an empty array for ready tasks, and then it extracts all elements of \textit{TaskPQ}, where their key is equal to the current time slot. This means that all predecessor tasks of these ready tasks have finished their execution. If \textit{TaskPQ} and \textit{ReadyTasks} array are both empty, the algorithm returns the final scheduling (\textit{Sch}) because it successfully schedules all tasks. 
If there is no ready task to be scheduled in the current time slot (the \textit{ReadyTask} array is empty, but the \textit{TaskPQ} is not empty), the algorithm moves to the next time slot (lines~2-10).

The algorithm sorts the ready tasks in descending order of their energy consumption (line 11). 
The energy consumption of each task $\tau_i$ ($Eng_i$) is calculated as follow:
\begin{align}
Eng_i=Pow_i\times~WCET_i  
\end{align}
where $Pow_i$, and $WCET_i$ are the maximum power consumption, and the worst case execution time of task $\tau_i$. 
The maximum power of each task can be obtained by running benchmarks on a real platform. As mentioned in Section~\ref{PowModel}, the processor power consists of three components; when a task is run on a processor, the dynamic power is increased significantly compared to static and independent powers. Hence, in this paper, we don't model the power; we measure the processor power when task are run on the real platform. More information about computing these values is given in Section~\ref{subsec:experimentalSetup}. \behnaz{The system's power consumption must never exceed the TDP constraint to overcome to overheating problem~\cite{Ansari2018,munawar14}. To this end, we consider a constant power~consumption for each task at design-time, which is equal to its maximum power consumption, to guarantee the meeting of TDP constraint in the worst-case scenario.}
In addition, energy increment leads to an increase in chip temperature~\cite{huang14,Ranjbar2020}. Thus, we map a task with more energy consumption to a core with less temperature. Then, the algorithm sorts the cores in the ascending order of their accumulated energy (line~12). A core has a higher priority for task assignment if it has~less accumulated energy (i.e., tends to have a less temperature degree).

\begin{figure*}

\centering

\subfloat[An example of MC application]{\includegraphics[width=0.33\linewidth]{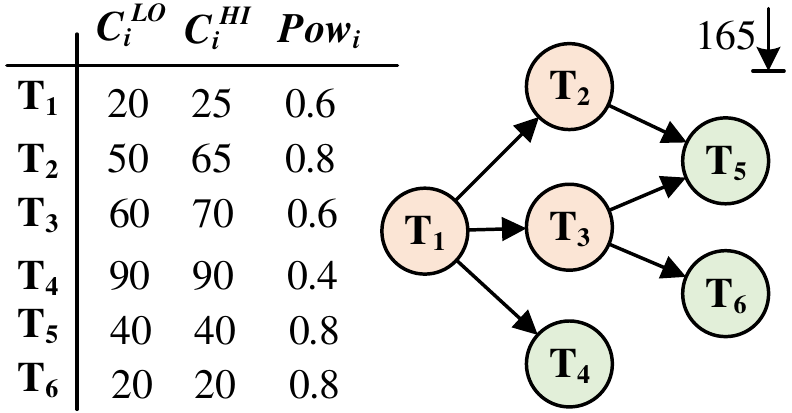}\label{fig: example application}}
\hspace{0.5cm}
\subfloat[Task scheduling without our policy]{\includegraphics[width=0.29\linewidth]{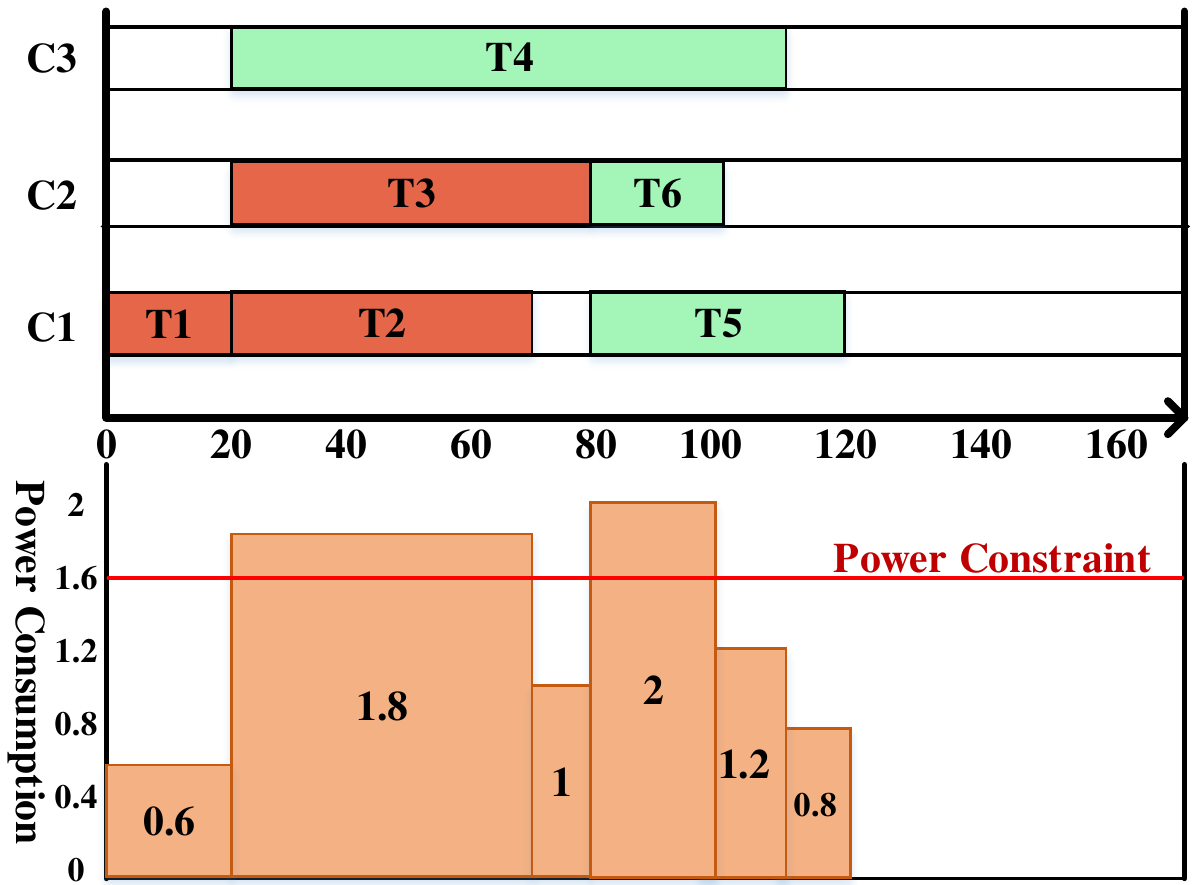}\label{fig:fig5b}}
\hspace{0.5cm}
\subfloat[Task scheduling with our policy]{\includegraphics[width=0.29\linewidth]{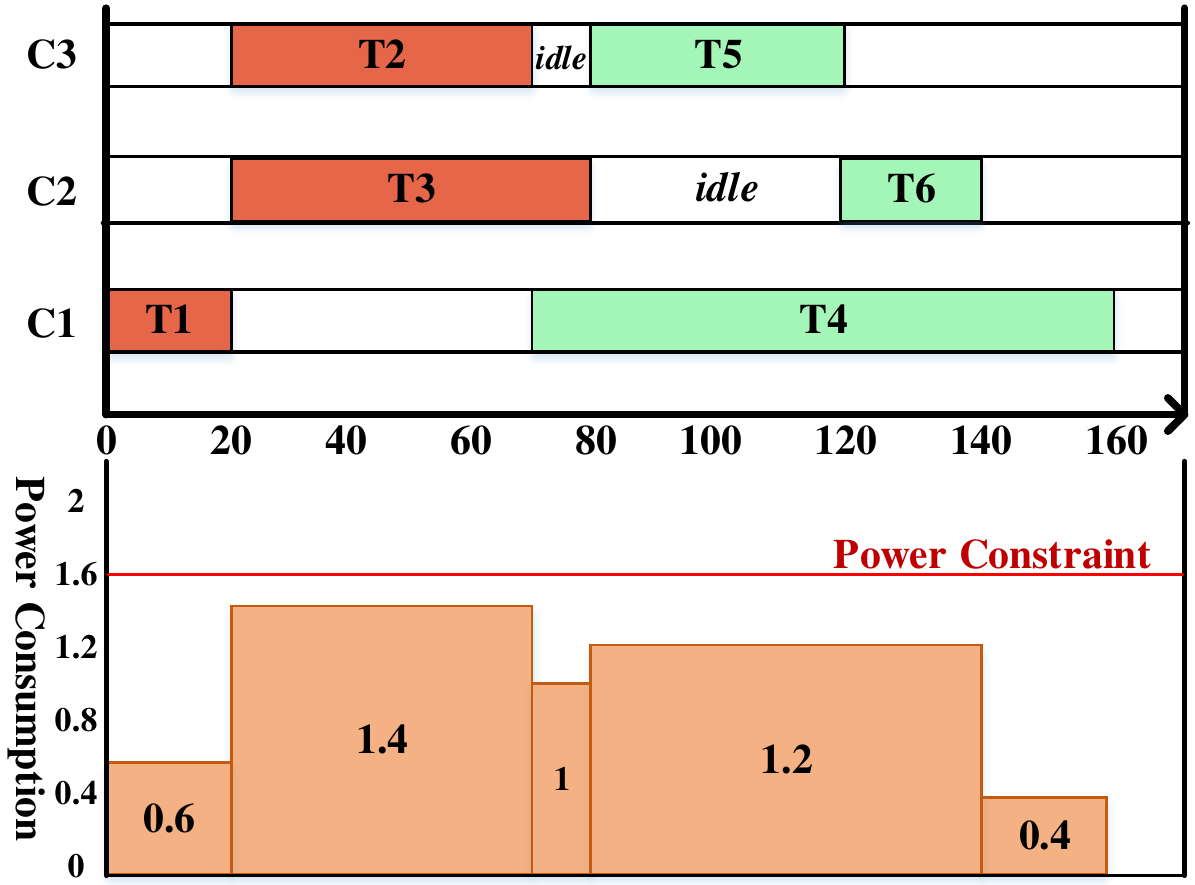}\label{fig:fig5c}}

\caption{Two different task scheduling scenarios}
\label{fig:scheduling5}
\end{figure*}

After sorting tasks and cores, the algorithm assigns tasks to the cores one by one (lines 13-38).
So, for each task, the algorithm selects a core from the sorted list and schedules~the task on the core's free slots (lines 19-26). 
The system's~instantaneous power consumption must be less than the TDP constraint; so, we consider an array called $SysPow$, which holds the maximum power consumption of the system in each time slot.
The algorithm checks the $SysPow$ and TDP constraint before scheduling a task on a core (line 20). If the task is completed before its deadline, the algorithm updates the schedule ($Sch$), power array ($SysPow$), and scheduling~status of the task ($task_{sch}$). It also sorts the cores again since the energy of one core has changed, and starts to schedule the next task (lines 27-33). If the task does not meet its deadline on the core, the algorithm picks the next core and schedule it on that core. However, if the deadline of one task is violated in all cores, the algorithm fails to schedule tasks of the application in this scenario and returns "un-schedulable" (lines 36-37). 


The example in Fig.~\ref{fig:scheduling5} shows an MC application and how our method maps and schedules the tasks on three cores. 
Assume the TDP constraint is 1.6 Watt. Fig.~\ref{fig:fig5b} shows that the scheduling without our policy violates the TDP constraint, while the maximum power consumption of the task scheduling by considering our policy is below the TDP constraint (Fig.~\ref{fig:fig5c}). 
When the system completes task $T_1$, three tasks ($T_2$, $T_3$ and $T_4$) become ready to be executed.  
So, $Eng_{T2}$ > $Eng_{T3}$ = $Eng_{T4}$ that shows $T_2$ should be mapped to the core with less accumulated energy ($E_c$). In addition, as can be computed, $E_{C3}$=$E_{C2}$>$E_{C1}$.
Therefore, according to the criticality level, we first map $T_2$ and $T_3$ on $C_3$ and $C_2$ and schedule them, and thereafter, $T_4$ is mapped on $C_4$. 
This procedure will be the same for mapping and scheduling $T_5$ and $T_6$.

It should be noted that the time complexity analysis and memory space analysis of the proposed method have been discussed in Appendix B and C, respectively.

\section{Experiment}
\label{sec:expriment}



\subsection{Experimental Setup}
\label{subsec:experimentalSetup}

\subsubsection{Application}
\label{subsec:appsetup}

\behnaz{For the experiments, we used both real-life and random applications to show our proposed approach's efficiency. To generate random task graphs, we used the tool presented by Medina et al. \cite{Medina18}.}
We generated applications with 30, 40, 50, and 100 tasks (\textit{n}), where 20\% to 50\% of them are LC tasks.
Another important parameter in a task graph is edge percentage (\textit{d}) that shows the probability of having edges from one task to other tasks. We considered 1\% to 20 \% edge percentage in the experiments.
Another important parameter that will be discussed in the experiments is 
the normalized system utilization $U/c$, where $U$ is the utilization of the system considering the high WCET of each task, and $c$ is the number of cores.
\behnaz{We considered different values of normalized system utilization in the range of (0,1] with the steps of 0.05. We also evaluate the proposed approach and other approaches for comparison, with a real-life application task graph, vehicle cruise controller (CC)~\cite{izosimov10}, composed of 32 tasks, where 34\% of them are LC tasks. \answersnd{In addition, the value of edge percentage for this CC application is 7\%.}}

\subsubsection{Hardware Platform}
\label{subsec:archsetup}
\behnaz{To evaluate our approach, we conduct the experiments and run the applications on a platform with 2, 4, 8, and 16 cores, which models ARM Cortex-A7 cores~(\textit{c}).}
The maximum number of transient faults that may occur during each application period (\textit{k}) and the recovery overhead~$\mu$ are considered three and 15ms, respectively~\cite{Izosimov08}.
\behnaz{It is important to know that if $\lambda$ and $time$ is the fault rate and application execution time, respectively, the minimum number of fault occurrence would be $\lambda\times time$. Therefore, $k$ would not be much smaller or larger than $\lambda\times time$~\cite{Zhang2006}. If $\lambda=10^{-6}fault/\mu s$, and $time=10^{3}ms$, then, $\lambda\times time=1$. As a result, since this fault rate is much higher than real fault rates, mentioned $10^{-12}fault/\mu s$ in~\cite{Salehi2016}, considering $k\leq 3$ is reasonable fault occurrence number during each application period.}
We use the HOTSPOT tool~\cite{huang06} to obtain the cores' temperature trace by exploiting the specific floorplan according to a real platform, ODROID XU3 board, which has four ARM Cortex-A7 cores, and the parameters used in \cite{Gong2018}.
In addition, we use the reported value in \cite{Chisholm17}, to consider the timing overhead of mode switching for ARM Cortex processors. 
We considered the maximum reported overhead, which is $T_{sw}= 254 \mu s$, in our experiments.

\subsubsection{Peak Power Consumption}
\label{subsec:experimentalSetuppowe}

\answersnd{To determine a realistic power consumption for tasks, we ran several embedded benchmarks from the MiBench suite, such as automotive, network, and Telecomm., on ARM Cortex-A7 core of ODROID XU3 platform with maximum frequency at design-time. We monitor the power sensors continuously, and we set the worst measured power as the power consumption of tasks.}
In addition, we examined different scenarios of activating one core to all cores by running different benchmarks. 
Each benchmark is run 1000 times on a core, and we considered each task's power consumption between the minimum and maximum power values obtained from the platform. The measurement reports show that the power consumption of tasks is between 483\textit{mW} to 939\textit{mW}.
In this paper, the TDP value has been considered 85\% of the maximum power that a chip can consume, which is used conventionally in embedded processors \cite{liu07}.

\subsubsection{Comparison}
We analyzed our proposed method and compared our experimental results to the results obtained by recent works that use the task graph model~\cite{socci15,Medina18,Ranjbar19}. Socci et al. \cite{socci15} have proposed an online scheduling algorithm for an MC system where only HC tasks are executed in the HI mode. Medina et al. \cite{Medina18} has considered a fault-tolerant MC system that generates two tables at design-time and uses them at run-time (see \sectionRef{sec:relatedWork} for more details). Based on this work, researchers in~\cite{Ranjbar19} have presented an online approach to reduce the peak power and temperature by using the DVFS technique. Hence, due to the timing overhead of DVFS and increasing fault rate by changing the \textit{V-f} levels~\cite{koren07}, we cannot easily use this technique especially in the HI mode. 

\subsection{Experimental Results}
\label{subsec:experimentalResults}

\subsubsection{\textbf{Tree Construction Time}}
\begin{figure}[t]
	\centering
	\begin{minipage}[t!]{1\linewidth}
		\centering
        \subfloat[Number of tasks]{\includegraphics[width=0.48\columnwidth]{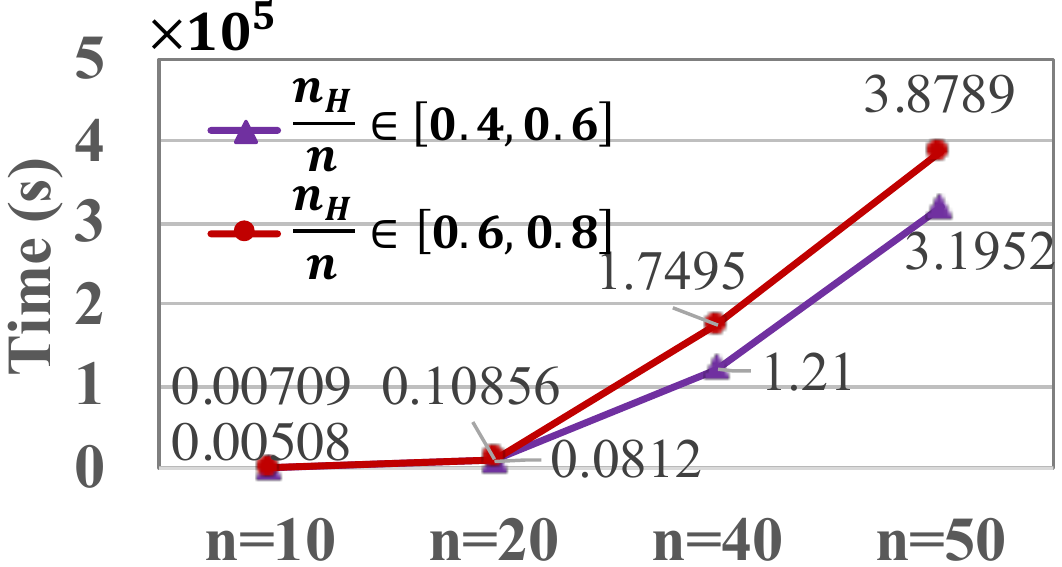}\label{VaryingTasksTime}}
        \hspace{4pt}
        \subfloat[Number of fault occurrence]{\includegraphics[width=0.48\columnwidth]{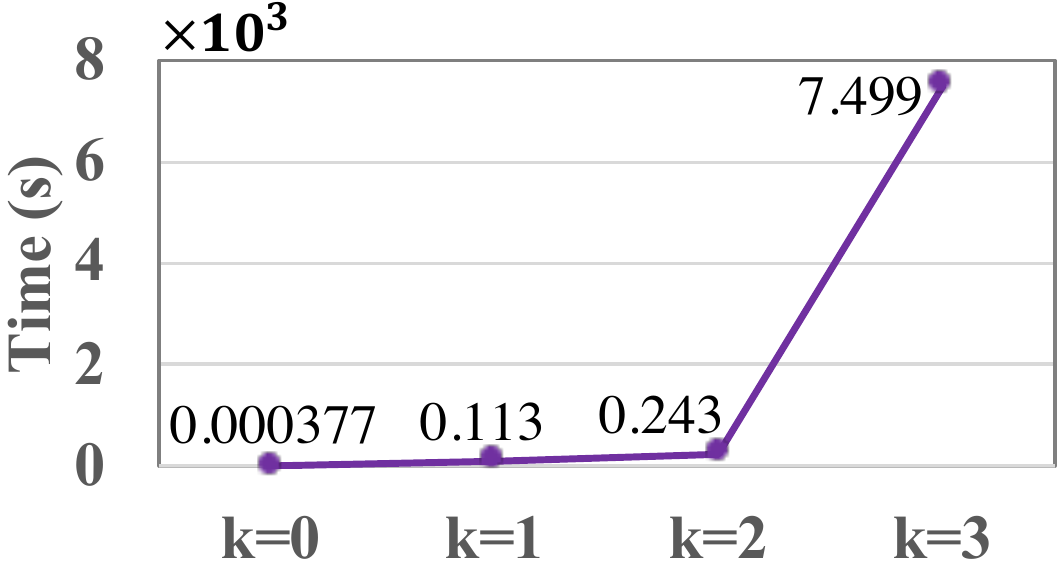}\label{VaryingFaultTime}}
		\caption{Tree construction time for different application sizes and number of fault occurrence}
		\label{fig: constructtime}
	\end{minipage}
\end{figure}

At first, we evaluate offline tree construction time by varying the parameters $n$ and $k$, in \figureRef{fig: constructtime}. 
The tree's construction time is computed on a system with an intel core-i5 processor with 1.3 GHz clock frequency. 
Construction time depends on the number of faults and tasks. \figureRef{VaryingTasksTime} shows the effects of the number of tasks and the portion of HC tasks in each task set with $k=3$. Besides, \figureRef{VaryingFaultTime} shows the effects of the number of fault occurrence with $n=20$. 
These figures depict that by increasing the number of faults or number of tasks, the tree generation's time is increased exponentially. Also, task sets with higher HC tasks have higher tree construction time. 
Although the offline tree construction time is relatively high for large applications, the online overhead is small and constant for all applications. It is noticeable that our method can generate each node of a tree in parallel to reduce the construction time. For example, if we have a system with four cores, the construction time is about four times faster than a single-core system. 

\subsubsection{\textbf{Run-Time Timing Overheads}}

In case of fault occurrence or mode switching, the system finds the proper schedule by moving to the child of the current node, which is responsible for the upcoming scenario. 
Each node of the tree stores two arrays with the size of $n\times (log^{c}_2+log^{\frac{period}{timeslot}}_2)$ bits, where $c$ and $n$ are the number cores and tasks, respectively.
Thus, the switching time between the schedules consists of moving one level in the tree and retrieving the correct scheduling from memory, which is constant and negligible.
We measured the schedule changing time at run-time on the ODROID XU3 platform, considering $c=8$, and $n=50$, it is almost 0.47$\mu$s.

%

%

\subsubsection{\textbf{Peak Power Management and Thermal Distribution for a Real-Life Application}}
\label{subsec:CCPowTempTrace}

\begin{figure}[t]
	\centering
	\includegraphics[width=0.765\columnwidth]{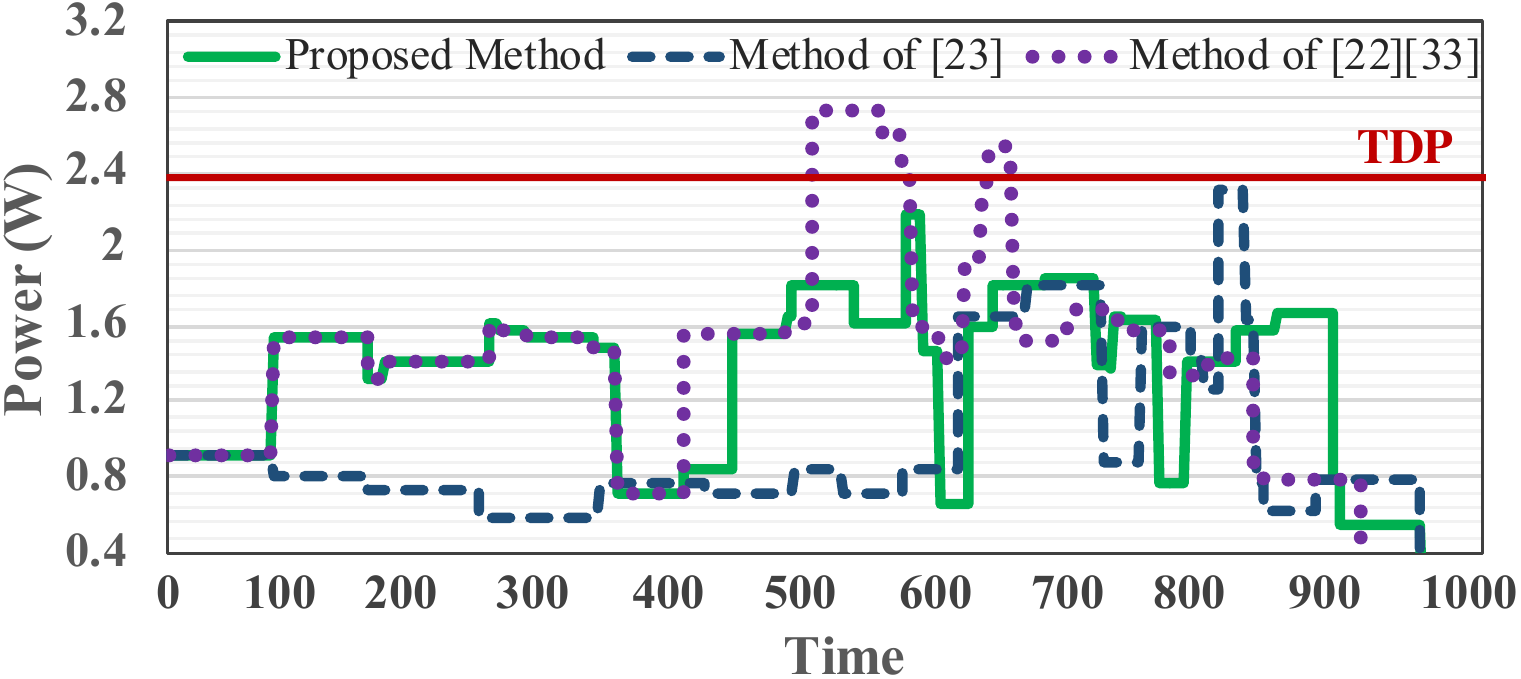}
	\caption{Power trace of real-life application graph (CC) in different methods under worst-case scenario}
	\label{fig: PowerCC}
\end{figure}

\behnaz{\figureRef{fig: PowerCC} shows the system's power traces of a real-life application, vehicle Cruise Controller (CC)~\cite{izosimov10} by our proposed approach, the approaches proposed by Socci et al.~\cite{socci15}, Medina et al.~\cite{Medina18}, and Ranjbar et al.~\cite{Ranjbar19}. Since~\cite{Ranjbar19} is the online approach to minimize the peak power, while exploiting the same task mapping and scheduling of~\cite{Medina18} at design-time, their power traces and thermal distributions are the same in the worst-case scenario of tasks' execution time and power consumption. 
}
In this part, to focus on the behavior of systems in the HI mode, we assumed no fault occurred during the application's execution.
Socci's approach does not violate TDP constraint because it drops all LC-tasks when the system switches to the HI mode, which means it has zero-percent QoS of LC tasks in the HI mode.
On the other hand, methods of~\cite{Medina18} and~\cite{Ranjbar19} guarantee 90.91\% of LC tasks execution in the HI mode, but it frequently violates the TDP constraint (\figureRef{fig: PowerCC}).
For the CC application, our method endeavored to execute 81.82\% of the LC tasks without violating TDP constraint. 
Appendix~D shows the thermal distribution of CC application and also, the corresponding results for a random task set with high utilization.


\behnaz{As a result, our method reduces the peak power and maximum temperature by up to 20.06\% and 3.71\% respectively, compared to the approach of~\cite{Medina18,Ranjbar19}, while the QoS is degraded 9.09\%. On the other hand, although our method increases the maximum temperature by 9.61\%, compared to~\cite{socci15}, we reduce the peak power consumption by 6.31\% and improve the QoS by 81.82\%.}

\subsubsection{\textbf{Analyzing the QoS of LC Tasks}}
\label{subsec:QoSAnalysis}

\begin{figure}[t]
	\centering
	\includegraphics[width=1\columnwidth]{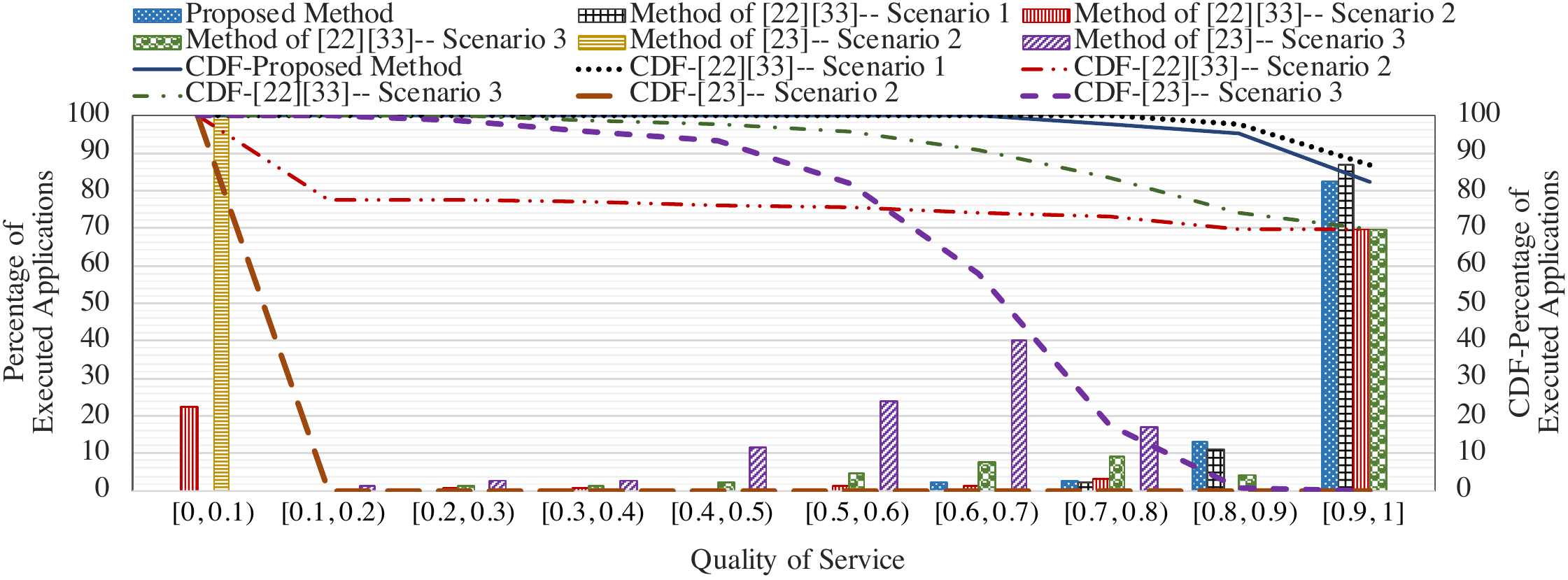}
	\caption{Task sets' QoS under different scenarios}
	\label{LCTQoS}
\end{figure}

\begin{figure*}[t!]
	\centering
	\begin{minipage}[t!]{1\linewidth}
		\centering
        \subfloat[Varying Utilization]{\includegraphics[width=0.25\columnwidth]{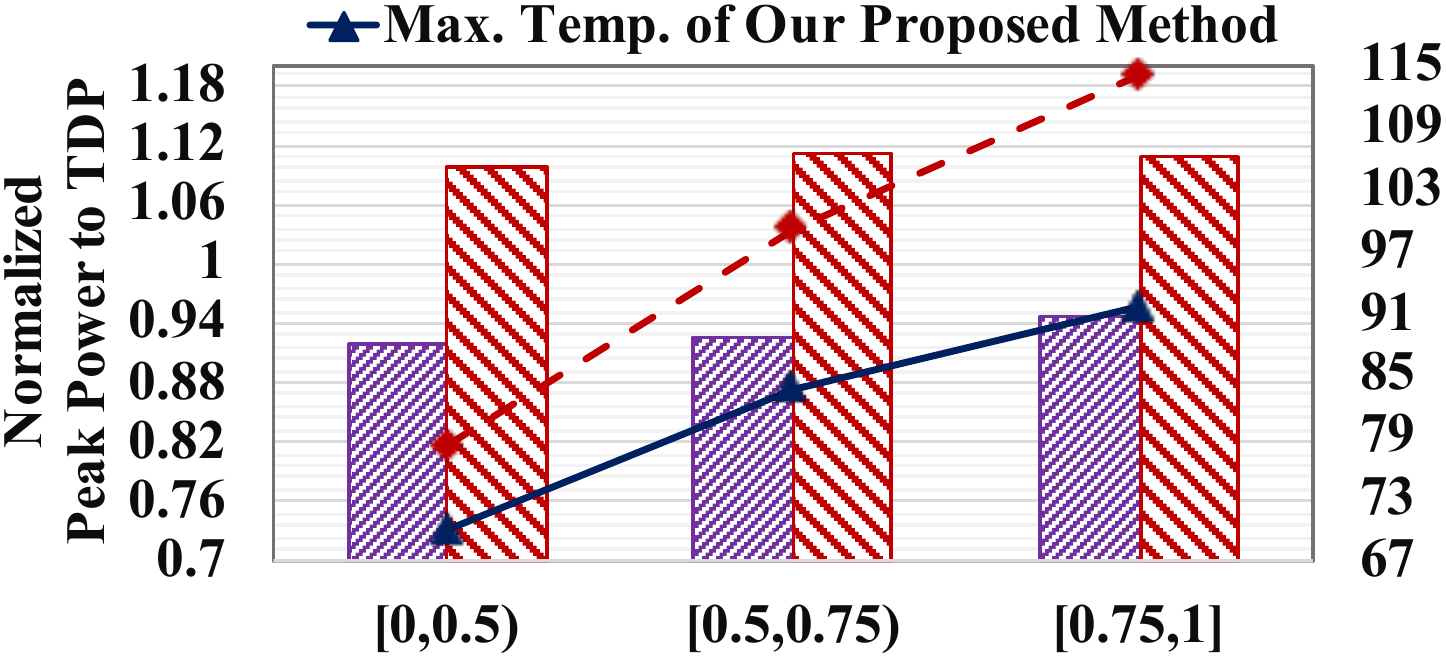}\label{PTVaryingUtil}}
        \hfil
        \subfloat[Varying Number of Tasks]{\includegraphics[width=0.23\columnwidth]{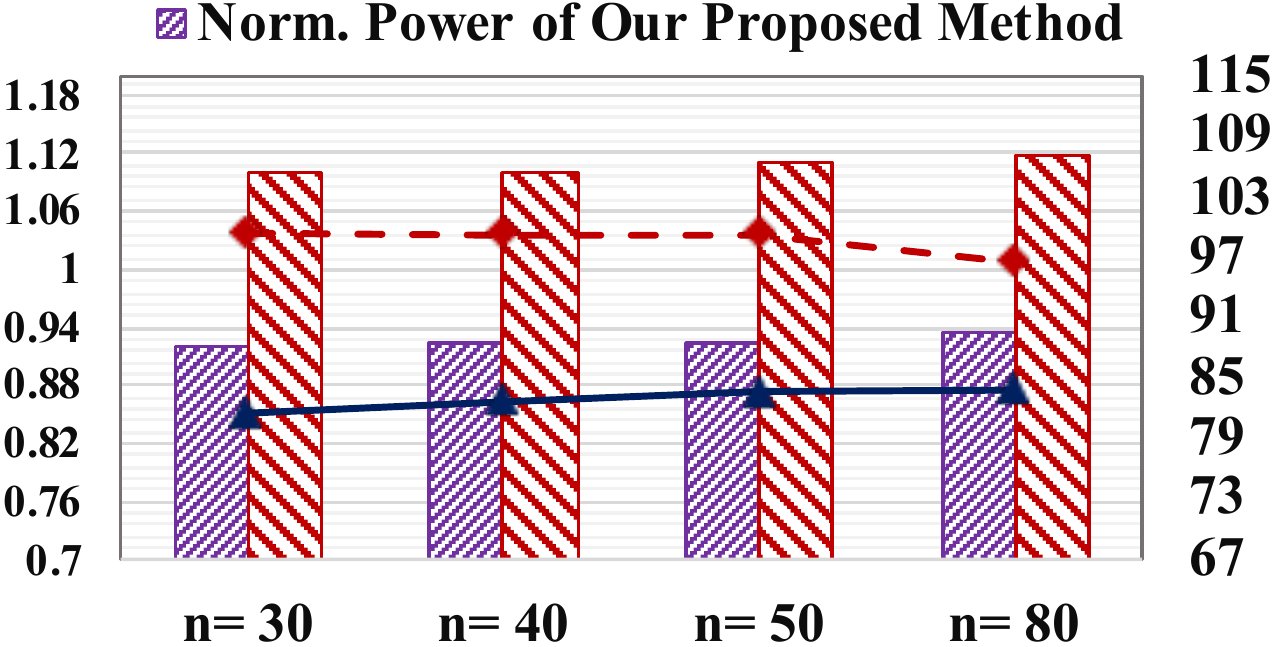}\label{PTVaryingTasks}}
        \hfil
        \subfloat[Varying Edge Percentage]{\includegraphics[width=0.23\columnwidth]{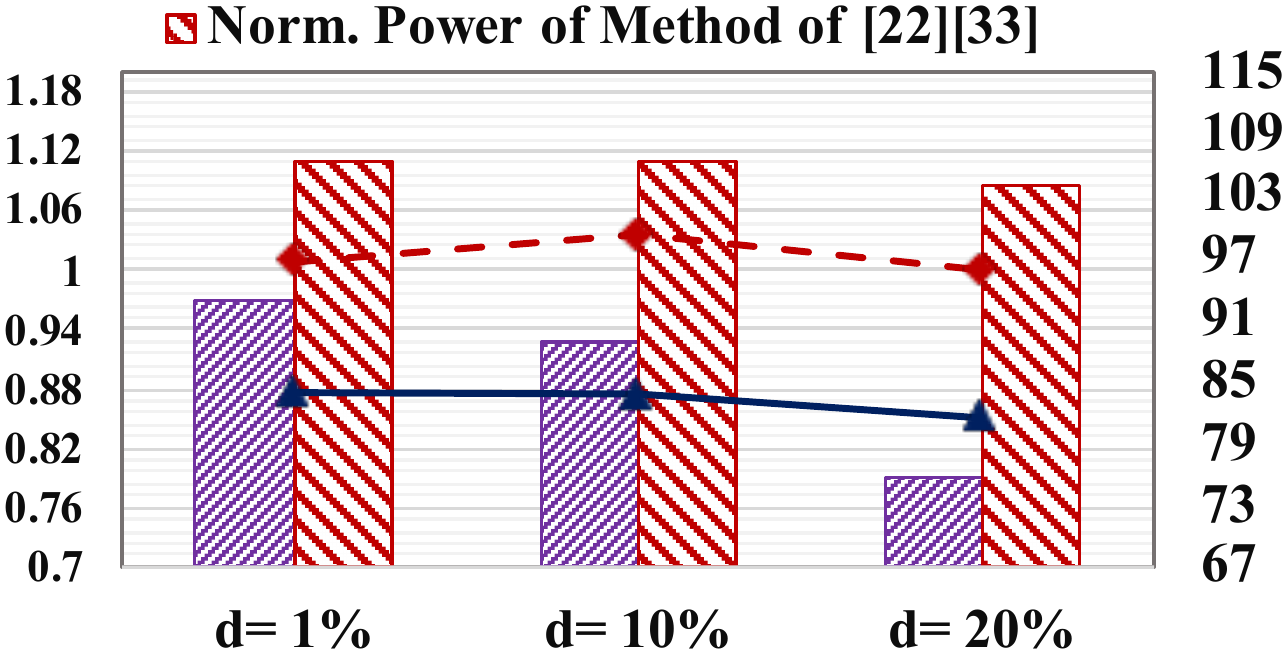}\label{PTVaryingEdge}}
        \hfil
        \subfloat[Varying Number of Cores]{\includegraphics[width=0.25\columnwidth]{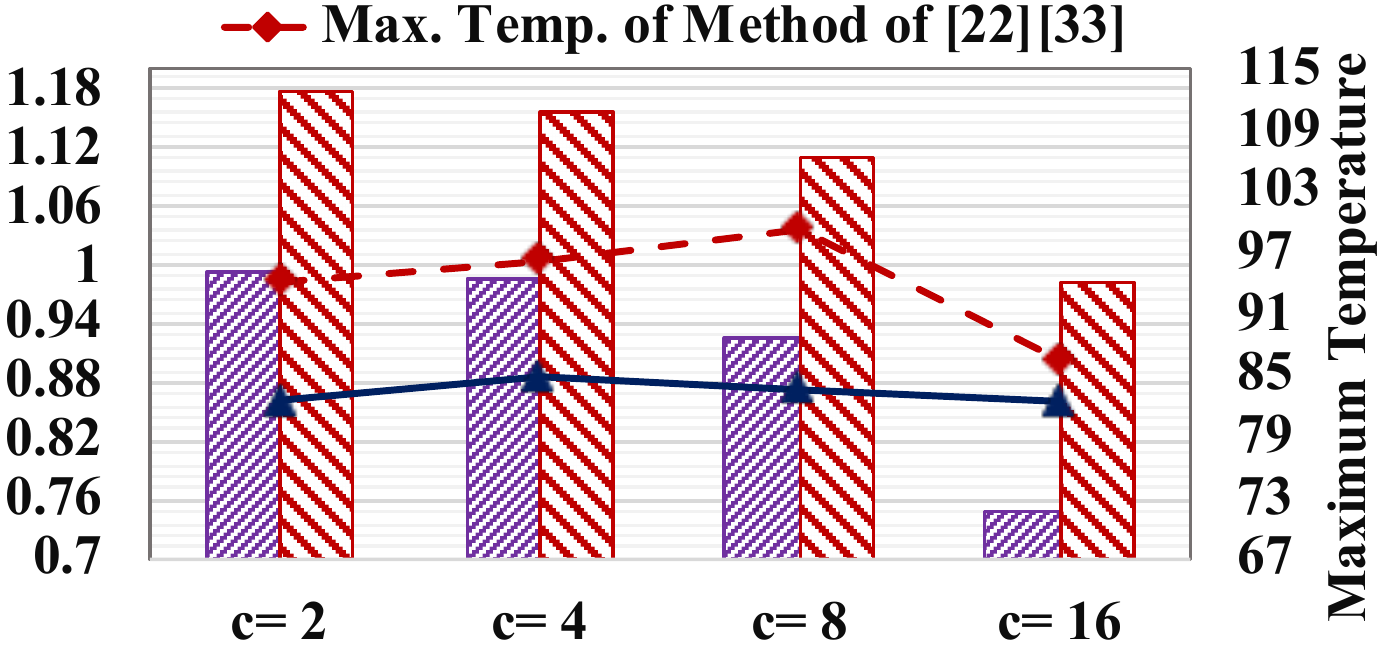}\label{PTVaryingCores}}
		\caption{Peak power consumption and maximum temperature in different methods}
		\label{PPowerResult}
	\end{minipage}
\end{figure*}

\behnaz{Now, we analyze the QoS for the proposed method in comparison with methods of~\cite{socci15,Medina18,Ranjbar19} in~\figureRef{LCTQoS}.} 
In our proposed method, there are many possible scenarios (each node of the tree is responsible for keeping the scheduling of system in one scenario), and for each scenario, the LC tasks' QoS is different. 
Thus, we run 100 schedulable task sets on eight cores, and for each task set, 10 different random situations of occurring faults and mode switching are considered. For the case when $U/c$=[0.5,0.75) or $U/c$=[0,0.5), there is more free slack before task set deadline; therefore, in the case of fault occurrence and mode switching, fewer LC tasks are dropped. 
In this experiment, we consider the worst-case scenario of processor demands, $U/c$= [0.75,1], $n$= 50 and $d$=10\% and in the generated task sets, 20\% to 50\% of tasks are LC tasks. In addition, the number of fault occurrences in each scenario is randomly selected in the range of [0,4]. \figureRef{LCTQoS} shows the LC tasks' QoS for all these 1000 scenarios. 
\behnaz{In Section~\ref{subsec:mixedCriticalityTaskModel}, we have defined the QoS as the successfully executed LC tasks to all LC tasks. However, we use three different definitions of QoS to evaluate the methods of~\cite{Medina18,Ranjbar19,socci15} more accurately, as follows. 
\begin{itemize}[leftmargin=*]
    \item Scenario~1: The QoS refers to How many LC tasks are successfully executed before their deadlines with no TDP violation. If TDP is going to be violated, running LC tasks are only stopped to reduce the peak power consumption.
    \item Scenario~2: This scenario has the same definition as Scenario~1, with the difference that since the HC tasks are the most important, therefore without them, QoS of LC tasks is penalized by completely being zero. Thus, if TDP is violated and some HC tasks are running on cores, then the QoS=0. 
    \item Scenario~3: Since those methods have not been specifically designed for peak power management while meeting the real-time constraints of all HC tasks, we give the HC tasks higher weighted and then consider the joint QoS, including both LC and HC tasks. Therefore, the HC tasks have a double weight in this scenario compared to LC tasks, in the case of dropping tasks due to the TDP violation.
\end{itemize}}

\behnaz{As shown in~\figureRef{LCTQoS}, the QoS for methods of~\cite{Medina18,Ranjbar19} in Scenario~1 is higher than the QoS for our proposed method in total (according to the Cumulative Distribution Function (CDF) line). However, in this scenario, the TDP constraint is violated several times due to the execution of HC tasks in parallel on cores, and there is no policy to manage the peak power. Moreover, for the second scenario in the methods of~\cite{Medina18,Ranjbar19}, the figure shows that in 22.5\% of task sets, the TDP constraint is violated while executing some HC tasks on cores. Besides, the QoS in the second scenario for~\cite{socci15} is zero, due to dropping all LC tasks in the HI mode. Now, if we investigate the methods of~\cite{Medina18,Ranjbar19,socci15} in the third scenario, the QoS of~\cite{Medina18,Ranjbar19} is more than the QoS of~\cite{socci15} due to the executing LC tasks in the HI mode. However, they have less QoS in this scenario compared to our proposed method. 
According to this figure, we can conclude that our proposed method is more efficient in improving QoS than other methods in different scenarios.} In addition, the minimum and maximum QoS in our proposed method are 68.33\% and 100\%, while the power constraint is always met.


\subsubsection{\textbf{Peak Power Consumption and Maximum Temperature Analysis}}
\label{subsec:PowTempAnalysis}

In this section, we evaluate the system's peak power consumption and the chip's maximum temperature in our approach and methods of~\cite{Medina18,Ranjbar19} by varying different parameters, such as utilization bound ($U/c$), number of tasks ($n$), edge percentage ($d$), and the number of cores ($c$). Fig.~\ref{PPowerResult} shows the average normalized peak power consumption to the TDP constraint and maximum temperature in the worst case that the system switches to the HI mode after executing the first HC task and all tasks execute up to their higher WCET. Hence, in the worst-case scenario, \cite{Ranjbar19} has the same power profile and thermal distribution as~\cite{Medina18}.

First, we analyze the peak power consumption by varying different parameters. The figure shows that our proposed approach can manage the peak power consumption to be less than the TDP constraint in all scenarios, while Medina's method violates the TDP constraint almost all scenarios. 
\behnaz{In general, the impact of our approach is increased as the probability of using parallelism in the execution of tasks on cores is increaseed (a large number of cores (larger \textit{c}) or less dependency between tasks (lower \textit{d})). 
Since the number of tasks and utilization are not changed by increasing \textit{c}, the maximum power consumption by~\cite{Medina18} is also reduced. However, since our proposed method endeavors to distribute the tasks on all cores to minimize hotspots and also minimize the instantaneous power consumption, our proposed approach in decreasing the peak power consumption is more efficient, compared to~\cite{Medina18}, while the \textit{c} is increased. Besides, by reducing the dependency between tasks (\textit{d}), although the system peak power consumption is increased and TDP is violated in~\cite{Medina18}, our approach always guarantees that the TDP constraint is not violated.}
In addition, although increasing the number of tasks or utilization increases the system's peak power consumption because the system does more computation, our approach guarantees that the TDP constraint will never be violated.

From the perspective of maximum temperature analysis, increasing the system utilization, illustrated in Fig.~\ref{PTVaryingUtil}, while the number of tasks is fixed ($n$=~50) means that the task's execution time tends to be longer. Thus, the computation time of cores is increased, the managing of peak power constraint and busy/idle times of cores would be difficult, and consequently, the maximum temperature of the chip is increased in both methods. However, \behnaz{we can decrease the maximum temperature by up to 
22.4$^\circ$C in comparison with~\cite{Medina18,Ranjbar19}. Besides, if we vary the number of tasks in Fig.~\ref{PTVaryingTasks}, since the computation time of cores does not change, the chip's maximum temperature is relatively constant by increasing the number of tasks in both our proposed method and methods of~\cite{Medina18,Ranjbar19}.} Additionally, by varying~\textit{d}, the maximum temperature reduces by increasing the dependency between tasks because the cores' computation time is constant, while the idle time of cores is increased. \behnaz{As shown in Fig.~\ref{PTVaryingEdge}, our proposed method can reduce the maximum temperature by 
14.3$^\circ$C on average by varying edge percentage, compared to~\cite{Medina18,Ranjbar19}.}

\begin{figure*}[t!]
	\centering
	\begin{minipage}[t!]{1\linewidth}
		\centering
        \subfloat[Varying Utilization]{\includegraphics[width=0.24\columnwidth]{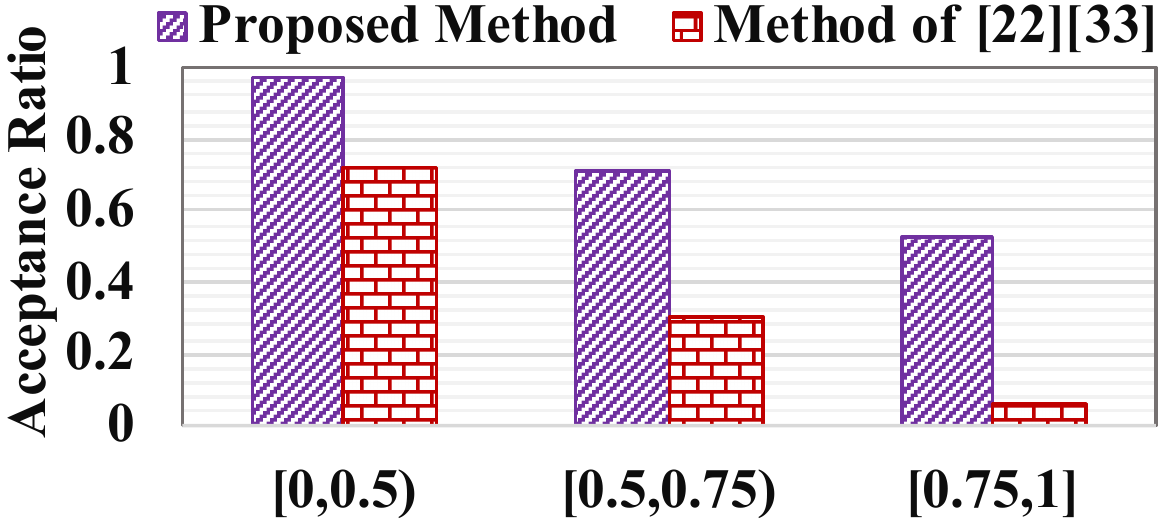}\label{VaryingUtil}}
        \hfil
        \subfloat[Varying Number of Tasks]{\includegraphics[width=0.23\columnwidth]{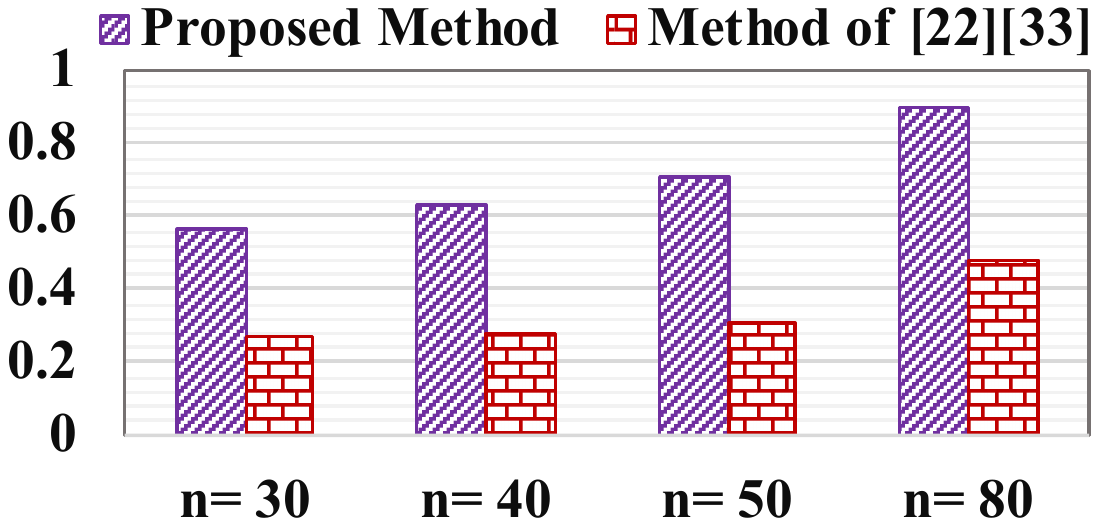}\label{VaryingTasks}}
        \hfil
        \subfloat[Varying Edge Percentage]{\includegraphics[width=0.23\columnwidth]{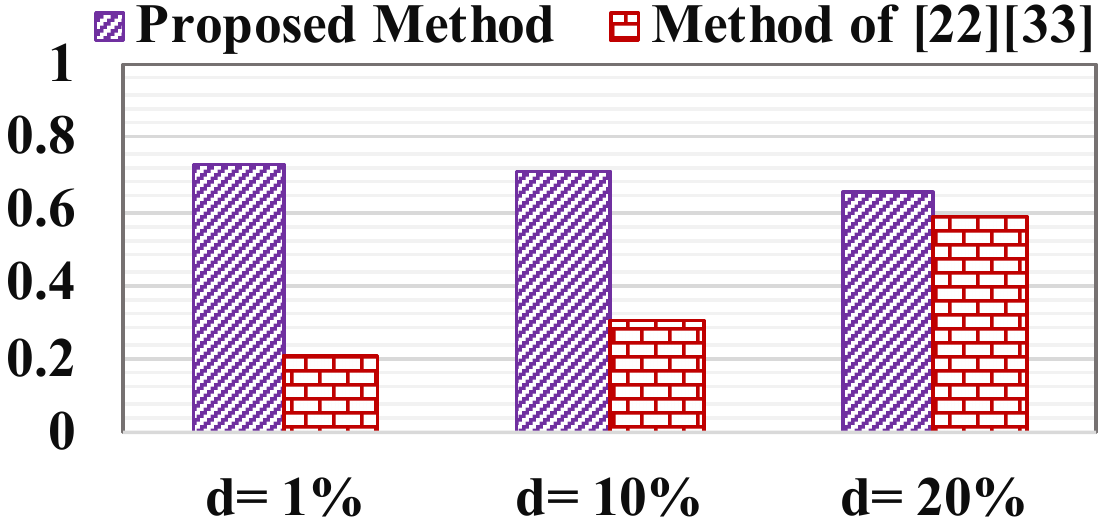}\label{VaryingEdge}}
        \hfil
        \subfloat[Varying Number of Cores]{\includegraphics[width=0.23\columnwidth]{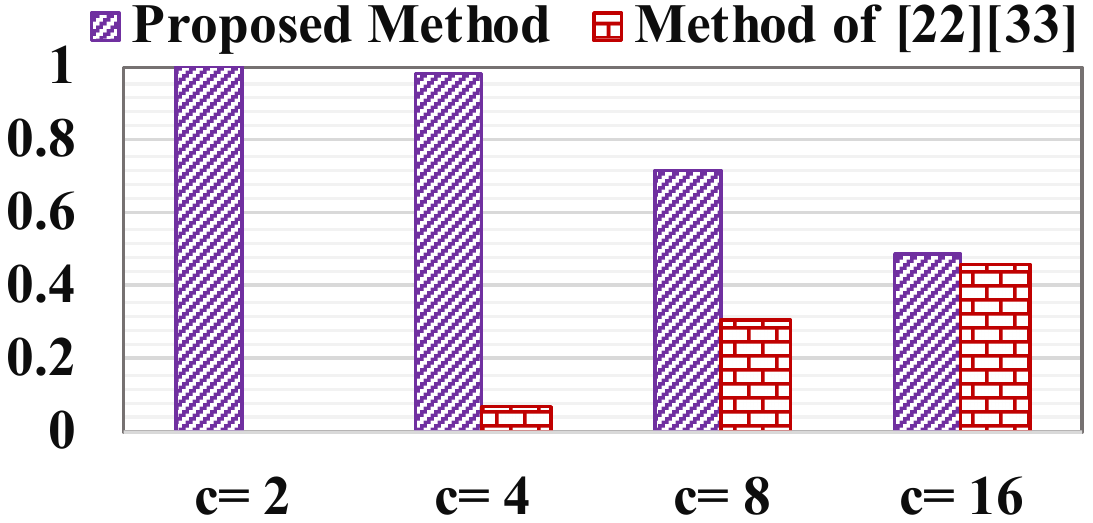}\label{VaryingCores}}
		\caption{Normalized acceptance ratio under different scenarios in different methods}
		\label{Varying}
	\end{minipage}
\end{figure*}

Now, we investigate the system's maximum temperature in Fig.~\ref{PTVaryingCores} by increasing the number of cores (\textit{c}) while other parameters are constant. Hence, the normalized utilization (\textit{U/c} is constant in this experiment, it means the utilization (U) is increased by increasing \textit{c}. We have more parallelism to execute tasks by increasing \textit{c} and add more free slack to let the cores be idle (having better thermal distribution). However, since each core's temperature is a function of its neighbor cores' temperature, it increases the chip's maximum temperature while the system utilization is increased. \behnaz{Therefore, the results of our proposed method show that the maximum temperature is relatively constant by increasing \textit{c} and can decrease it, 
10.7$^\circ$C on average, compared to~\cite{Medina18,Ranjbar19}. Since the methods of~\cite{Medina18,Ranjbar19} do not consider the thermal distribution, the maximum temperature is generally increased by increasing~\textit{c}.}

\subsubsection{\textbf{Effect of Varying Different Parameters on Acceptance Ratio}}
\label{subsec:SchAnalysis}

In this section, we illustrate the impact of different parameters, such as utilization bound ($U/c$), number of tasks ($n$), edge percentage ($d$), and the number of cores ($c$) on~the task schedulability (acceptance ratio). 
\behnaz{\figureRef{Varying} represents the effect of each parameter, while the others are fixed, to analyze how the proposed method and the methods of~\cite{Medina18,Ranjbar19} react to each parameter. 
We run 1000 benchmarks for each scenario, 
and report the average result. A task set is schedulable if the real-time and power constraints are met.
} 
In general, having more dependency between tasks, large system utilization, or more cores, causes the system to have less acceptance ratio in our proposed method. We discuss the observation in detail.

From the perspective of utilization bound, we fix other parameters to see the effective of varying utilization in~\figureRef{VaryingUtil}. Increasing the utilization while the number of tasks is fixed ($n$=~50), means that the tasks execution time tends to be longer. when the utilization is getting higher, the computation time of cores is increased. Therefore, fewer task sets can be scheduled before their deadlines even in the case of fault occurrence, and also, the managing of power constraint and busy/idle times of cores would be difficult. 
We can conclude that fewer task sets can be scheduled before their deadline, while the power constraint is not violated. \behnaz{This trend is also the same with~\cite{Medina18,Ranjbar19} with the difference that the TDP constraint is violated several times, which causes the task sets not to be schedulable.}

Besides, \figureRef{VaryingTasks} shows that the task schedulability is increased by increasing the number of tasks, with $d$=~10\%, $U/c$=~[0.5,0.75), and $c$=~8. Since the system utilization is constant for all number of tasks, the execution time of tasks is reduced by increasing the number of tasks. Therefore, the tasks tend to finish their execution early and allow their successors to be released. Furthermore, the overhead of re-executing a task due to fault occurrence is much lower for small tasks.
According to~\figureRef{VaryingTasks}, we conclude that our method can schedule \behnaz{70\%} of task sets, on average, when only the number of tasks is varied. 
\behnaz{Besides, by increasing the number of tasks in the methods of~\cite{Medina18,Ranjbar19}, 
the parallel task execution is increased, which causes more peak power consumption, and therefore, less task schedulability due to the TDP violation. }



For the case of varying the edge percentage, when the dependency between the tasks is increased, while $n$, $U/c$ and $c$ are constant ($n$=~50, $U/c$=~[0.5,0.75), $c$=~8), the release time of tasks is increased, because tasks must wait for more predecessor tasks to finish their executions. 
Therefore, the idle time on cores increases, which causes more delays in the execution of tasks, and reduce the schedulability. 
\figureRef{VaryingEdge} shows that the highest schedulability (73\%) is achieved in our method when $d$=1\%. 
\figureRef{VaryingCores} shows the effect of varying the number of cores in the system on task schedulability when other parameters are not changed ($n$=~50, $U/c$=~[0.5,0.75), $d$=~10\%). 
By considering the fixed $U/c$, the utilization is increased by increasing the number of cores.
Consequently, the execution time of tasks is increased  
because the number of tasks is fixed. 
As mentioned earlier, the task schedulability is decreased by increasing the tasks' execution time. Therefore, as can be seen in~\figureRef{VaryingCores}, the schedulability of applications with our proposed method decreases by increasing the number of cores, while the other parameters are fixed. 
Besides, in methods of~\cite{Medina18,Ranjbar19}, the acceptance ratio increases by increasing the dependency between tasks and having more cores in the system. The reason is that based on their mapping and scheduling algorithm, \answersnd{by increasing the edge percentage and number of cores while fixing other parameters, tasks have less parallelism, and also fewer cores are selected to be active to execute the tasks, which causes the system to have less peak power consumption and therefore, a higher acceptance ratio. As a result, by increasing the edge percentage for more than 20\%, our proposed method and method of~\cite{Medina18,Ranjbar19} have almost the same acceptance ratio.} However, the mapping and scheduling algorithm of~\cite{Medina18,Ranjbar19} increases the overheating problem, which is not acceptable by most safety-critical systems. 
In addition, since methods of~\cite{Medina18,Ranjbar19} are not peak-power aware, 
when the number of cores is less, the TDP is violated in most of the task sets and therefore, these task sets are not schedulable.

\behnaz{In the end, the acceptance ratio of our proposed method is 74.14\% on average for all scenarios, while it is 31.1\% in the methods of~\cite{Medina18,Ranjbar19}.}

\subsubsection{\textbf{\behnaz{Investigating Different Approaches at Run-Time}}}
\label{subsec:RuntimePowTrace}

\begin{figure}[t]
	\centering
	\includegraphics[width=0.95\columnwidth]{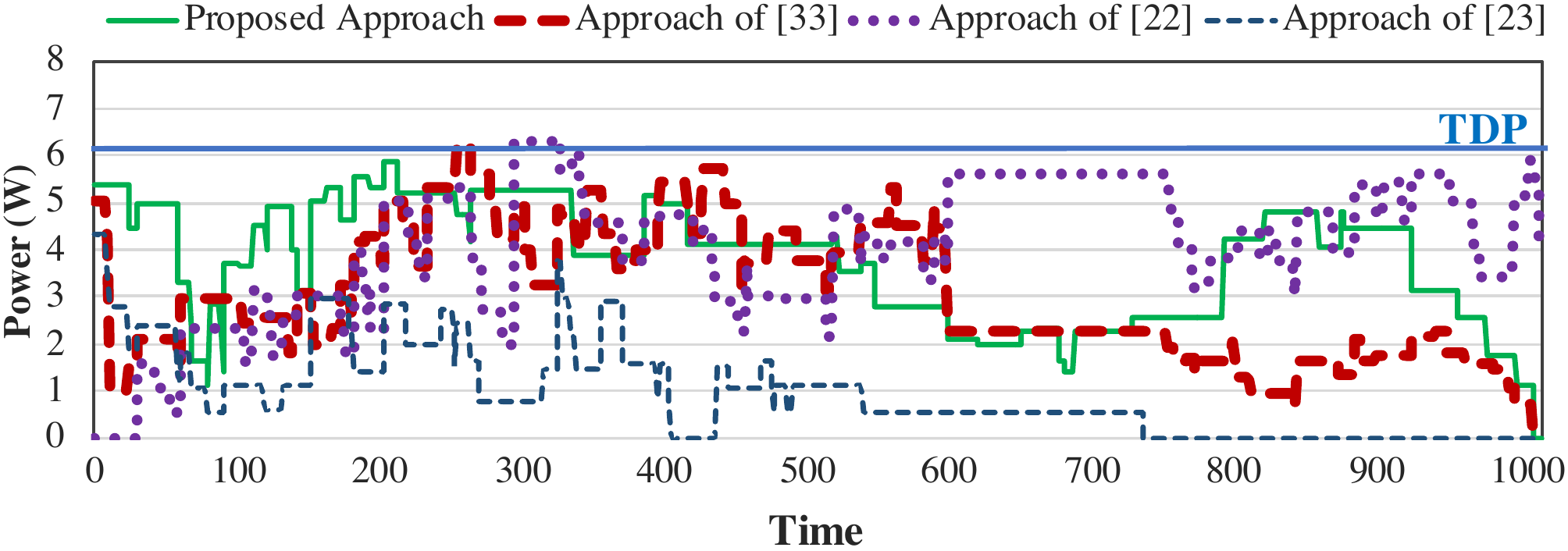}
	\caption{System power trace of different methods at run-time}
	\label{Runtimepowertrace}
\end{figure}

\behnaz{Now, we evaluate the system behavior at run-time in terms of peak power consumption for our proposed approach and the method proposed in~\cite{Ranjbar19,socci15,Medina18}.} Researchers in~\cite{Ranjbar19} have presented a run-time method to reclaim the available slacks and reduce the \textit{V-f} levels of cores to decrease the system peak power. Here, analogous to~\cite{Ranjbar19}, the actual execution time of tasks follows the normal distribution with the mean and standard deviation of $\frac{3\times WCET}{4}$ and $\frac{WCET}{12}$, respectively. Fig.~\ref{Runtimepowertrace} depicts the run-time power trace of methods for a random task graph with $d=10\%$, $n=50$, $c=8$ and $U/c=0.9$. The system switches to the HI mode by forcing a randomly selected HC task to execute beyond its lowest WCET for both methods. \behnaz{As shown in this figure, the system peak power in the proposed approach is less than the TDP constraint at run-time, while the method of~\cite{Medina18} has violated the TDP for a period of time. Although the system peak power of~\cite{Ranjbar19} may be less than the TDP constraint for some applications like the used task graph and their method consumes less energy in the system, there is no guarantee for the peak power to be less than the power constraint. Due to the using of DVFS technique in~\cite{Ranjbar19} and decreasing the \textit{V-f} levels at run-time, the system consumes less energy in comparison with our proposed method. For the example of Fig.~\ref{Runtimepowertrace}, the method of~\cite{Ranjbar19} saves 0.74102$J$ in system compared to our proposed approach.}
However, the DVFS technique degrades the reliability and increases the fault rate. The fault rate depends on the system's voltage level, and also, the application's reliability depends on the voltage level and tasks' WCET, which is increased by reducing the frequency level~\cite{Salehi2016}. 
As an example, for this task graph, by considering the fault rate $f=10^{-4}$~\cite{salehi16}, the reliability of tasks has been decreased $0.17\%$ and $2.08\%$, on average and worst-case in comparison with our proposed approach. In addition, the number of nines for the system's reliability ($-Log_{10}^{1-Rel}$) has been degraded from 8 to 6, which may not be desirable for most safety-critical applications~\cite{Ranjbar2020Fantom}.

From the perspective of system's reliability and fault-tolerance in our proposed method, we run the 1000 task graph applications for different normalized utilization bound ($U/c$= [0.5,0.75), and [0.75,1]) and compare to the system's reliability in~\cite{Ranjbar19}. In our proposed method, $-Log_{10}^{1-Rel}$ for the normalized utilization equal to [0.5,0.75), and [0.75,1], is 8.56 and 7.67 on average, respectively, while for ~\cite{Ranjbar19}, is 4.99, and 4.60, respectively. 
As a result, based on the required reliability for safety-critical systems, the method of~\cite{Ranjbar19}, which decreases the \textit{V-f} levels, has severely damaged the system's safety, which is not desirable. The reliability of the proposed approach is high for different utilization. Therefore, our method can be applied to any application with varying bounds of utilization while satisfying the peak power management, fault-tolerance, and high reliability in different system operational modes. 

\behnaz{Besides, we have evaluated the proposed method and compared with the other methods at run-time by using the real task graph in Appendix E.}

\section{Conclusion}
\label{sec:conclusion}
This paper has proposed an approach to schedule MC tasks in fault-tolerant systems in different operational modes to manage peak power by considering a thermal management policy. 
At run-time, depending on the fault occurrence possibilities and criticality mode changes, the system faces different scenarios.
We proposed an approach that develops a tree of schedules at design-time. Each node of the tree represents a scenario and contains scheduling where the system is able to execute all HC tasks and as many as possible LC tasks without violating the TDP.
At run-time, a low overhead online scheduler selects the proper node to map and schedule tasks. The results show that the proposed technique can schedule 74.14\% of task sets on average and significantly reduce peak power consumption (by guaranteeing the TDP constraint) in the worst-case scenario compared to the existing methods. 

\behnaz{As a future work, we would design the MC systems~to~generate the scheduling tree based on tasks' average execution times instead of WCETs to improve the QoS in the HI mode.
}

\ifCLASSOPTIONcaptionsoff
  \newpage
\fi

\bibliographystyle{IEEEtran}
\bibliography{Reference.bib}



\begin{IEEEbiography}[{\includegraphics[width=1in,height=1.25in,clip]{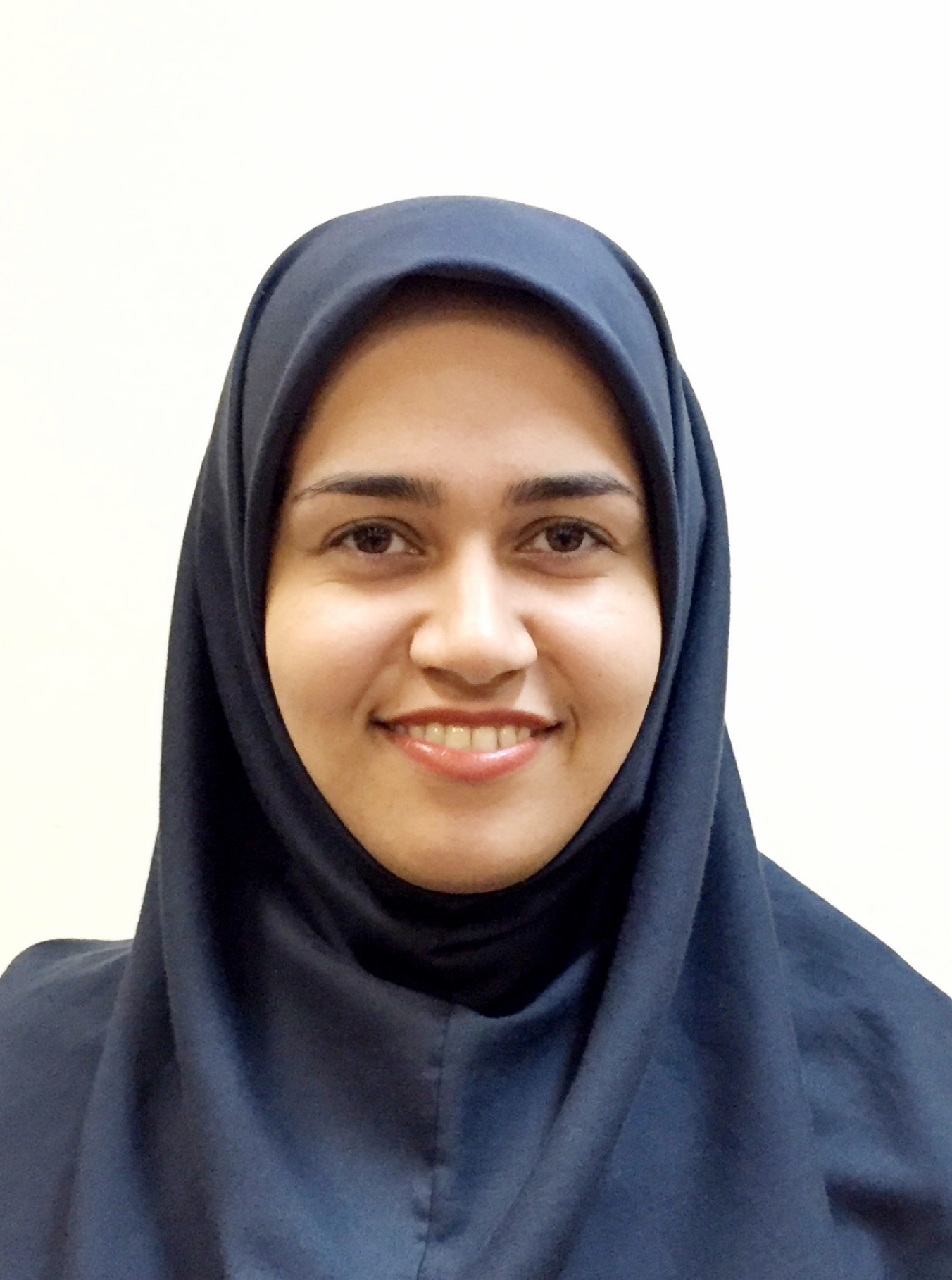}}]{Behnaz Ranjbar} received the B.S. degree in computer engineering from Amirkabir University of Technology in 2012 and the M.S. degree from Sharif University of Technology, Tehran, Iran in 2014. She is currently a joint Ph.D. student with the Sharif University of Technology, Tehran, Iran, and the Chair for Processor Design, Technische Universität Dresden, Dresden, Germany. Her research interest includes real-time, fault tolerant and low-power embedded system design.
\end{IEEEbiography}

\begin{IEEEbiography}[{\includegraphics[width=1in,height=1.25in,clip]{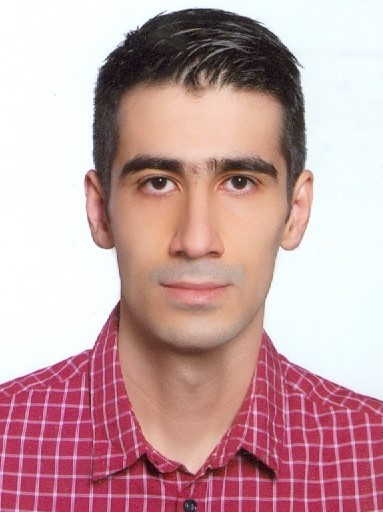}}]
	{Ali Hosseinghorban} received his B.Sc. in computer engineering from Shahid Beheshti University and his M.Sc. from Sharif University of Technology in 2015 and 2017, respectively. He is currently a Ph.D. student in the Computer Engineering department of Sharif University of Technology. He is also a visiting researcher at the Chair for Processor Design, Technische Universit{\"a}t Dresden, Dresden, Germany. His research interests include low-power real-time embedded systems, reconfigurable computing, and management of nonvolatile memories.
\end{IEEEbiography}

\begin{IEEEbiography}[{\includegraphics[width=1in,height=1.25in,clip]{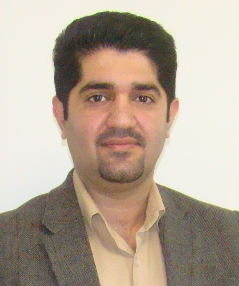}}]
	{Mohammad Salehi} received the PhD degree in computer engineering from Sharif University of Technology, Tehran, Iran, in 2016. He is currently an assistant professor of computer engineering at University of Guilan, Rasht, Iran. From 2014 to 2015, he was a visiting researcher in the Chair for Embedded Systems, Karlsruhe Institute of Technology, Germany. His research interests include design of low-power, reliable and real-time embedded systems with a focus on dependability and energy efﬁciency in cyber-physical systems.
\end{IEEEbiography}

\begin{IEEEbiography}[{\includegraphics[width=1in,height=1.25in,clip]{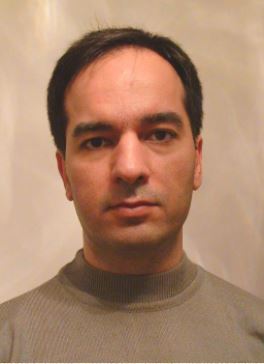}}]{Alireza Ejlali} received the PhD degree in computer engineering from Sharif University of Technology (SUT), Tehran, Iran, in 2006. He is currently an associate professor of computer engineering at SUT. From 2005 to 2006, he was a visiting researcher in the Electronic Systems Design Group, University of Southampton, Southampton, United Kingdom. He is currently the Director of the Embedded Systems Research Laboratory, SUT. His research interests include low power design, real-time systems, and fault-tolerant embedded systems.
\end{IEEEbiography}

\begin{IEEEbiography}[{\includegraphics[width=0.92in,height=1.25in,clip]{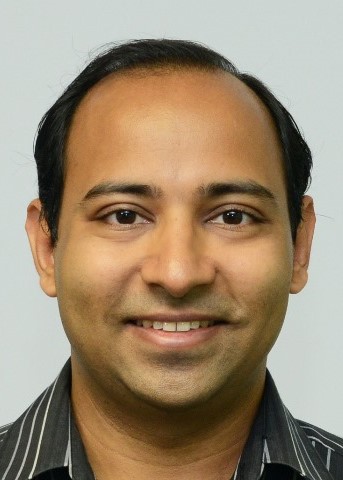}}]{Akash Kumar} (SM’13) received the joint Ph.D. degree in electrical engineering and embedded systems from the Eindhoven University of Technology, Eindhoven, The Netherlands, and the National University of Singapore (NUS), Singapore, in 2009. From 2009 to 2015, he was with NUS. He is currently a Professor with Technische Universität Dresden, Dresden, Germany, where he is directing the Chair for Processor Design. His current research interests include the design, analysis, and resource management of low-power and fault-tolerant embedded multiprocessor systems. 
\end{IEEEbiography}

\appendices

\section{A Brief Overview of Fault Detection and Correction Mechanisms}

Researchers in~\cite{koren07,siewiorek2017reliable,Bolchini13,Cerrolaza2020} give a comprehensive study in the field of fault detection and correction mechanisms. A comparison of recent mechanisms has been studied recently in~\cite{Thati2018}. In general, the key to these mechanisms is using redundancy, which could be information, hardware, time, or analytical. In the following, we present these mechanisms briefly, which have been used in recent works in the field of embedded real-time systems. 

\begin{itemize}
    \item Hardware Redundancy: Hardware redundancy is one of the most useful mechanisms to detect and correct a fault in embedded real-time systems, that can be classified as active, passive, and hybrid~\cite{siewiorek2017reliable,koren07}. Duplication is the most straightforward fault detection active redundancy, where two copies of a task are executed, and in the case of fault occurrence, the results of the two copies are not identical. However, this technique cannot tolerate faults. Therefore, N-modular redundancy with a voter and replication are passive redundancy and can be used to detect and correct the faults in applications, used recently in~\cite{Ansari2018,salehi16}. Watchdog timer and standby sparing are the other hardware redundancy techniques that standby sparing has been used in~\cite{Safari2020} recently. 

    \item Timing Redundancy: Check-pointing and rollback-recovery is one of the most common techniques to detect transient faults in embedded systems due to its cost-effectiveness~\cite{Salehi2016}. When faults are detected, the re-executed technique can be used to tolerate it. Finding the optimum number of check-points during the execution of tasks is one of the challenges when this technique is used.

    \item Information Redundancy: This redundancy is a software-based technique in which error detection codes or parity bits can detect and recover the faulty data. Some recent works like~\cite{Kajmakovic2019,asghari2020} have focused on mitigating the soft errors by using the information redundancy.
    \item Analytical Redundancy: A few recent research works~\cite{Iqbal2019,lee2020fault,park2018lired} have focused on detecting fault based on deep learning. The reason for using Machine Learning (ML) techniques is that some faults may not be detected by existing methods, which are applied at design-time and employed at run-time, and then a failure in identifying the error may have serious consequences. The ML mechanisms are classified into two categories of data driven-based, such as unsupervised learning, and knowledge driven-based such as reinforcement learning. Since the embedded MC systems are safety-critical, detecting faults must have low latency to not impact the tasks' deadlines. In addition, since these systems are embedded, the chosen ML technique must be low-power. In general, any learning technique has some advantages and disadvantages. Therefore, the designer can select the proper technique based on the feature of the systems. Additionally, researchers in~\cite{angelopoulos2020} have studied ML-based fault detection, prediction, and prevention and presented a survey of ML solutions.
\end{itemize}

\section{Time Complexity Analysis}
\label{subsubsec:treegenerate}

In this section, we consider an \textit{m}-core processor running an application with \textit{n} tasks, and \textit{k} possible fault occurrences during the application's execution. At first, we describe the complexity of generating a tree as the main computation part of the proposed method.

When an HC task overruns, the system switches to the HI mode, and the scheduler considers the high WCET of remaining tasks until complete execution of the application. So, in each execution, only one task may overrun, and the possible scenarios for overrun situations are equal to the number of HC tasks (\textit{$n_H$}).
Furthermore, we assume that up to \textit{k} faults may occur during the execution of the application. For clarity, we first compute the number of different fault occurrence scenarios for \textit{k} = 0, 1 and 2. Then, we present a general formula to obtain the maximum number of possible scenarios (nodes of the tree).

	\begin{itemize}[leftmargin=*]
	\item \textit{k} = 0:
	In this case, there is one scenario for a situation where none of the tasks overruns (the root of the tree) and $n_H$~(number of HC tasks in the graph) scenarios for situations where one of the HC tasks overruns (leaf nodes of the tree). Thus, the number of all schedules is:
	\begin{align}
	&T(k=0)=1+n_H
	\end{align}

    \item $\textit{k}\leq1$:
    In this case, the tree has $T(k=0)$ nodes to handle $k=0$ scenarios, in addition to the nodes which contain scheduling for $k=1$ scenarios.
    There are three possible situations for the scenarios where one fault occurs to a task (HC or LC task). There are $n$ scenarios for situations when no HC task overruns, $n \times n_H$ scenarios for situations where the faulty task executes before an HC task overruns, and $n\times n_H$ scenarios situations where the faulty task executes after an HC task overruns. Therefore:
    \begin{align}
    	T(k\leq1)&= T(k=0) + n + 2\times n\times n_H \nonumber\\
    	&=1 + n_H + n_H \times n + n\times (1+n_H) \nonumber\\
    	&=1 + n_H + n_H \times n + n\times T(k=0) 
	\end{align}
    
    \item $\textit{k}\leq2$:
    In this case, the tree has $T(k\leq1)$ nodes to handle $k\leq1$ scenarios, in addition to the nodes which contain scheduling for $k=2$ scenarios.
    There are four possible situations for the scenarios where two faults occur to one or two task(s). There are $n^2$ scenarios for situations when no HC task overruns, $n \times n \times n_H$ scenarios for situations where the two faulty tasks are executed before an HC task overruns, $n_H \times n \times n$ scenarios for situations where the two faulty tasks are executed after an HC task overruns, 
    and $n\times n_H \times n$ scenarios situations where an HC task overruns in the middle of two faulty tasks. It is noteworthy to mention that there are scenarios in which both faults and overrun happens on a one HC task (similar to S7, S10, S12 and S14 in Table~\ref{tab:Motivational}). Therefore:
    \begin{align}
	&T(k\leq2)=T(k\leq1)+n^2 + 3 \times n_H \times n^2 \nonumber \\ 
	&=1 +  n_H + n_H \times n + n_H \times n^2 + n\times T(k\leq1)
	\end{align}

    Therefore, we can conclude that the maximum number of possible schedules by considering maximum \textit{k} fault occurrence is:
	\begin{align}
		\label{complexity}
	    T(k)&= 1+n_H(\sum_{i=0}^{k}(n^{i}))+nT(k-1), 
	    T(0)=1+n_H
    \end{align}
    \end{itemize}

By solving the Eq.~\ref{complexity}, we can conclude that generating the tree is in order of \textit{$O(n^{k+2})$}.
\begin{align}
\label{eq:treetimec}
&T(k)= n_H\times(\sum_{i=0}^{k}((i+1)\times n^{i}))+\frac{1-n^{k+1}}{1-n}
\end{align}

Please note that this value is a general upper bound for the generating tree algorithm, and for an actual task graph, the total number of scenarios is less than Eq.~\ref{complexity}. The reason is that, the total number of scenarios are presented with no awareness of the exact dependency between tasks to count the precise number of scenarios when a fault occurs and then a task overruns, or vice versa. 
For example, there are 14 different scenarios for two HC tasks, one LC task, and k=1 for the task graph presented in~\figureRef{fig:a_taskGraph}, while $T(k)$ is equal to 18 in Eq.~\ref{complexity}.


\section{Memory Space Analysis}
\label{subsec:Memory Space Overhead}

In this section, we discuss the memory space needed for storing the scheduling tree. For each scenario, we store two arrays with the size of the number of tasks. The first array determines the core assigned to each task, and the second array determines the start time of the tasks. In the first array, we denote that each task is mapped to which of $c$ cores. So, each task requires $\log_2^c$ bits. 
Since we have \textit{n} tasks in the application, the total memory space required for each scenario is $n\times\log_2^c$ bits. 
Considering the period of the application and the size of each time slot, the second array size is equal to $n\times\log_2^{period/timeslot}$. Therefore, the total amount of needed memory (bits) is:
\begin{align}
    Mem(n,c,k) = T(k) \times(n\times(\log^c_2+\log^{\frac{period}{timeslot}}_2)
\end{align}

Assuming \textit{c} and \textit{period/timeslot} values are less than $2^{32}$, the memory space needed for saving the scheduling tree for an application with 32 tasks and up to two possible fault occurrences in the worst-case scenario is less than 13 MB.
\behnaz{It is noteworthy to mention that the scheduling tree can be stored in the FLASH or read-only memory of the system, and there is no need to load the whole tree to the RAM at run-time. In the case of fault occurrence or mode switching at run-time, the system discards the current schedule and loads the proper child node's schedule into the RAM. In this example, our approach occupies less than 2 KB of the RAM.}

\section{Power Trace and Thermal Distributions of the Real Application and a Random Task Graph Example under Worst-Case Scenario}
\label{A-ExpPowTemp}

This paper focuses on peak power management and maximum temperature reduction in fault-tolerance MC systems in multi-core platforms. We have evaluated different methods for a real-life task graph running on three cores in Section~\ref{subsec:CCPowTempTrace}. In this appendix, we first show the thermal distribution of the real-life application, and then, we show the corresponding results for a random application with high utilization.

\begin{figure}[t]
	\begin{minipage}[t]{1\linewidth}
		\centering
		\subfloat[Proposed Method]{\includegraphics[width=0.28\textwidth]{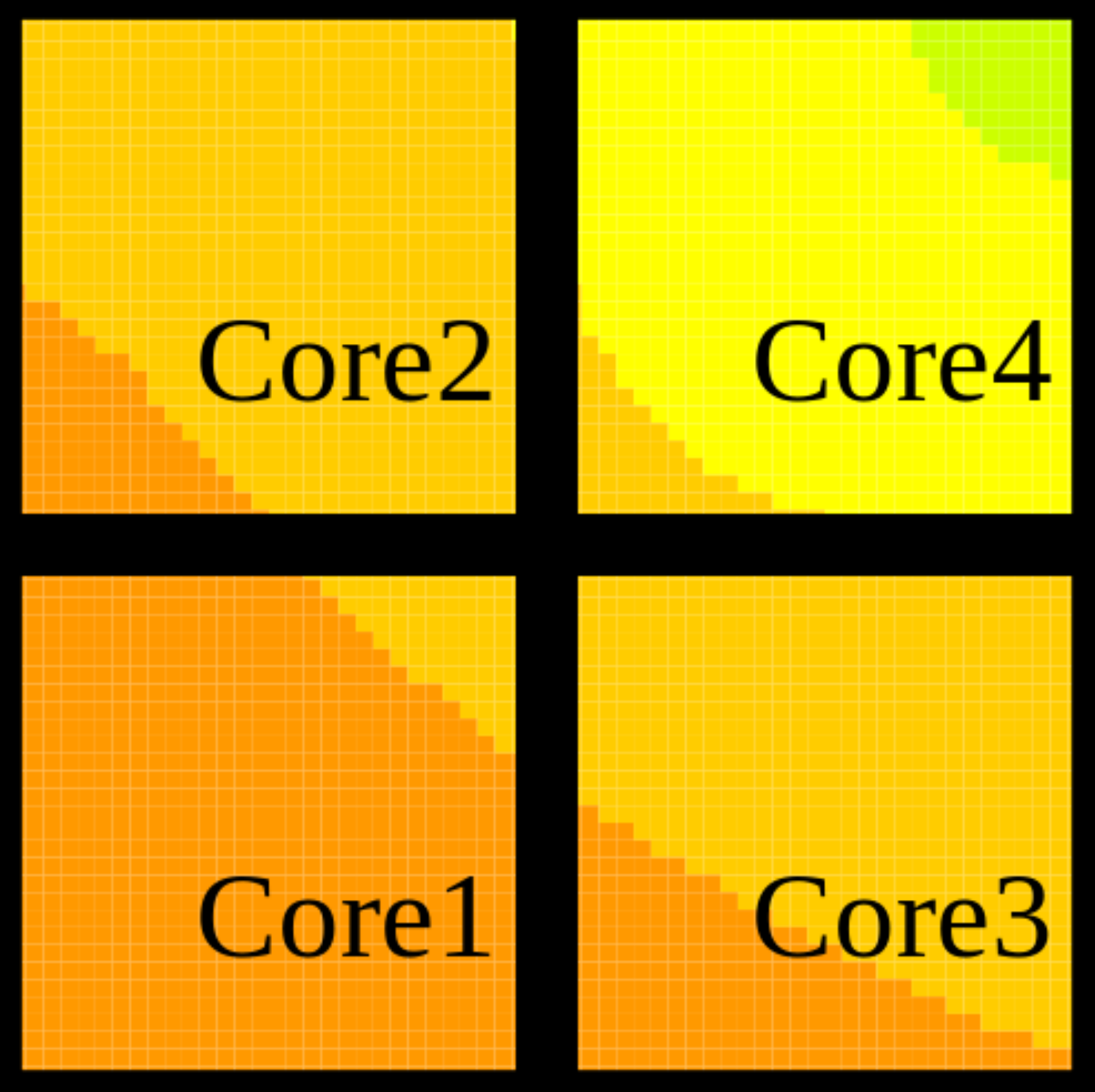}\label{fig:fig11a}}
        \hfil
        \subfloat[\cite{Medina18,Ranjbar19}]{\includegraphics[width=0.28\textwidth]{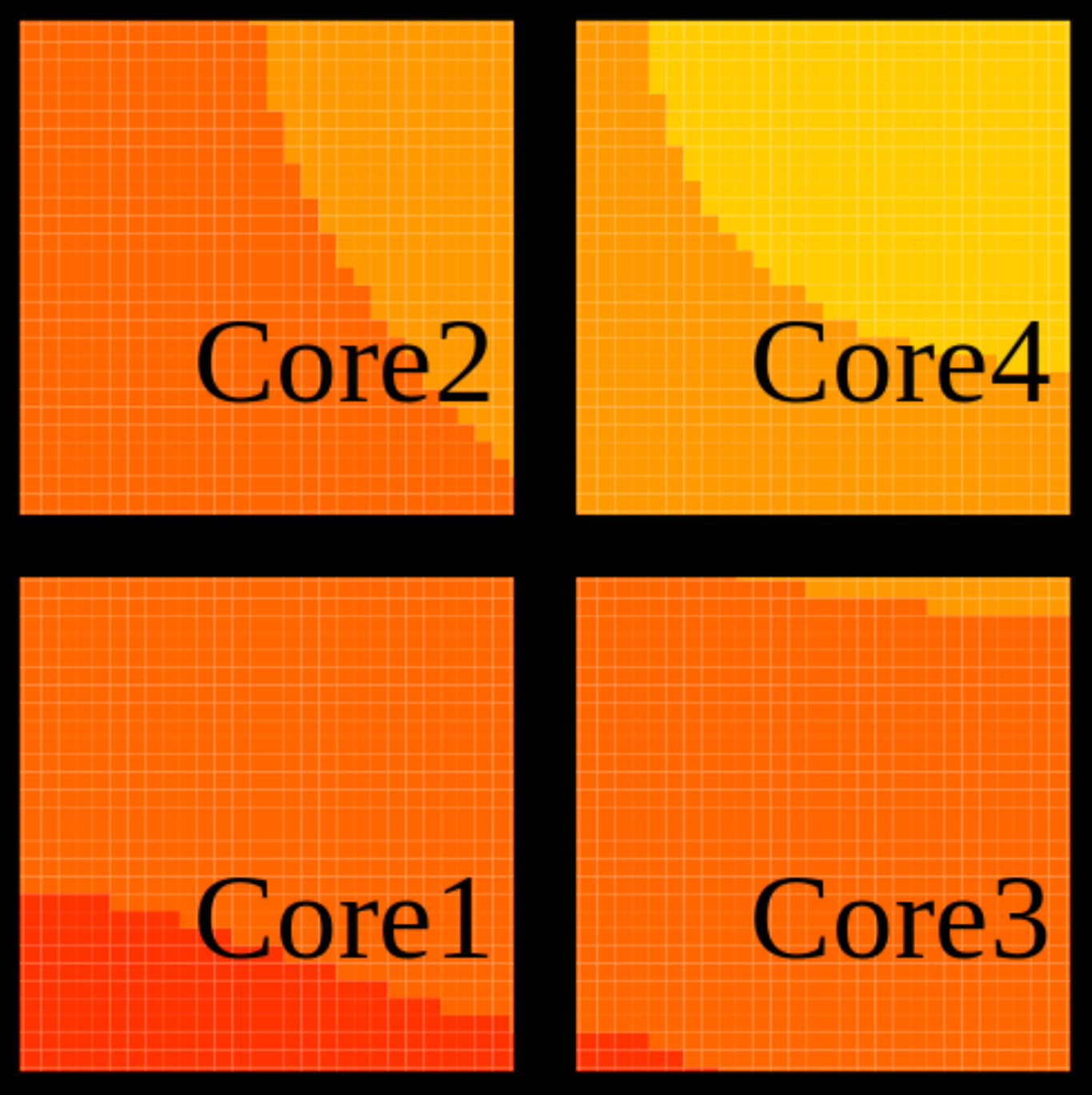}\label{fig:fig11b}}
        \hfil
        \subfloat[\cite{socci15}]{\includegraphics[width=0.28\textwidth]{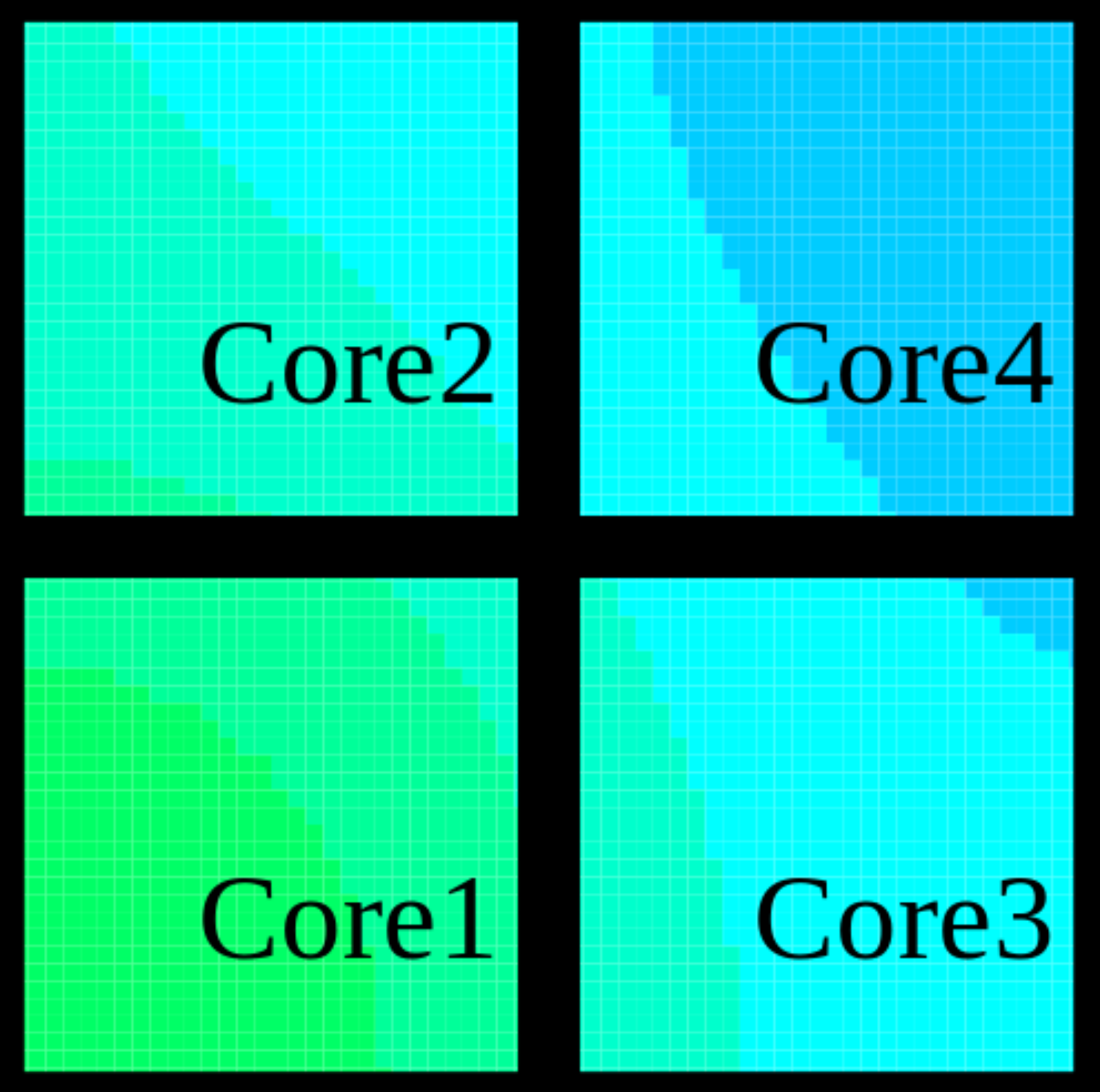}\label{fig:fig11c}}
        \hfil
        \subfloat{\includegraphics[width=0.07\textwidth]{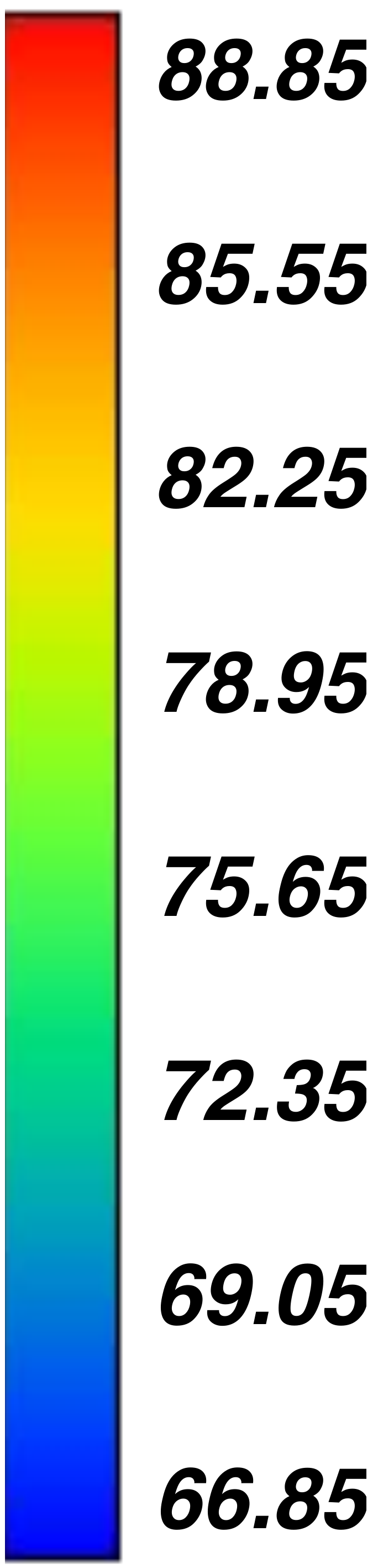}\label{fig:fig11d}}
		\caption{Thermal profiles of real-life application graph (CC) in different methods under worst-case scenario}
        \label{fig:realLifeScenarios}
	\end{minipage}
\end{figure}

\figureRef{fig:realLifeScenarios} shows the steady-state temperature distribution of Sooci~\cite{socci15}, Medina~\cite{Medina18}, Ranjbar~\cite{Ranjbar19}, and our proposed method using the HOTSPOT simulator for the CC application, corresponding to power profile of Fig.~7. Although the maximum temperature of~\cite{socci15} is lower than ours, it has zero-percent LC tasks' QoS because it does not execute any LC task in the HI mode. 
In addition, we map the tasks on the cores more uniform than the Medina's method, which prevents hotspots in our approach. The proposed approach could reduce 5$^\circ$C in maximum temperature compared to~\cite{Medina18,Ranjbar19}. If the system becomes larger in terms of number of cores and tasks, the efficiency of our proposed approach in reducing the hotspots would be higher.

\begin{figure}
	\centering
	\includegraphics[width=1\columnwidth]{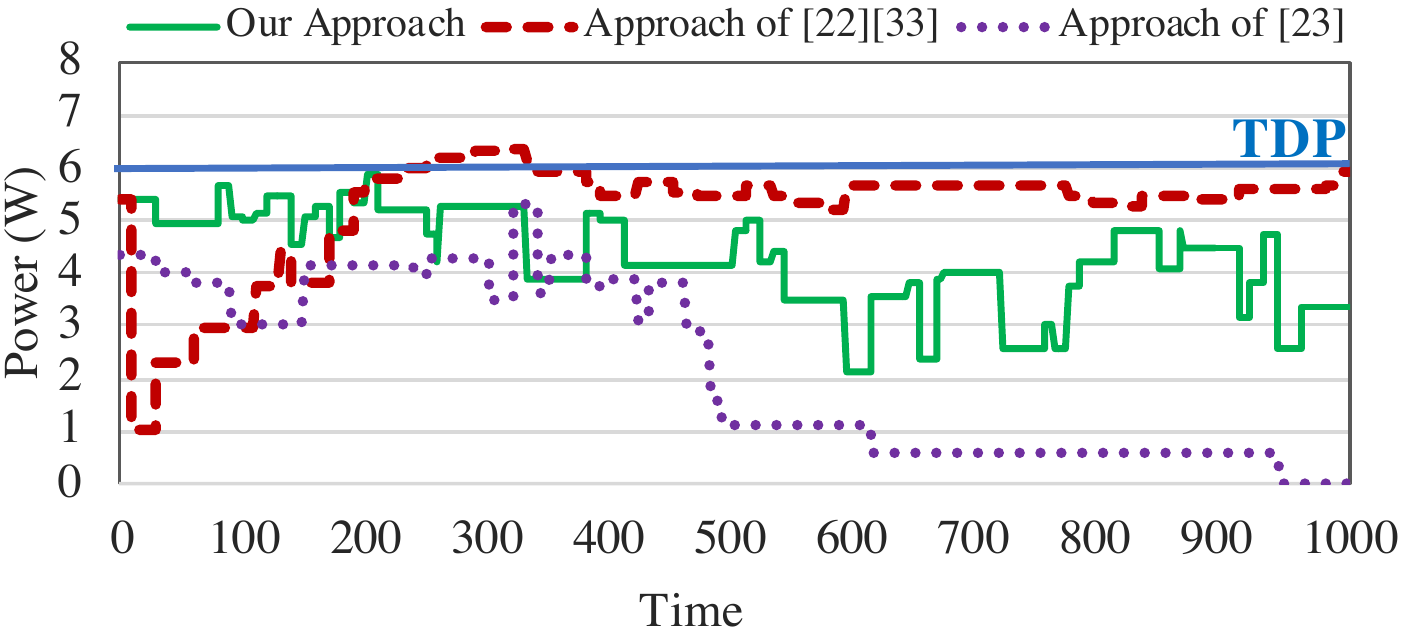}
	\caption{Power trace of a random task graph in different methods under worst-case scenario}
	\label{fig: Powerexp}
\end{figure}

\begin{figure}
	\begin{minipage}[t]{1\linewidth}
		\centering
		\subfloat[Proposed Method]{\includegraphics[width=0.28\textwidth]{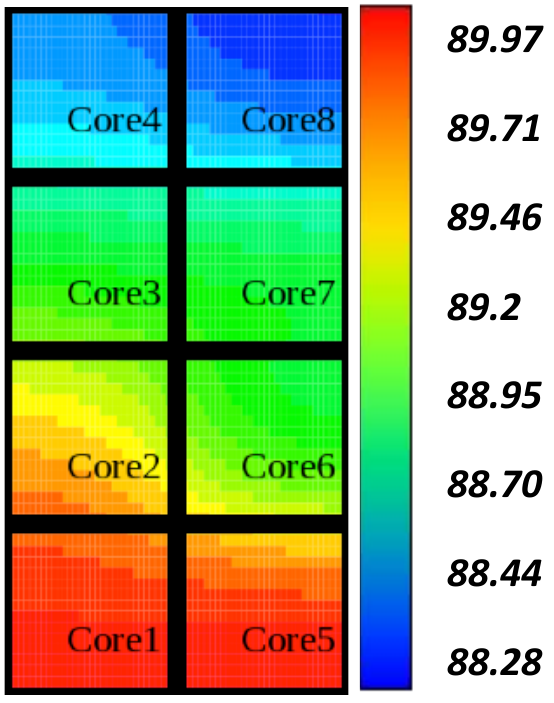}\label{fig:fig14a}}
        \hfil
        \subfloat[\cite{Medina18,Ranjbar19}]{\includegraphics[width=0.29\textwidth]{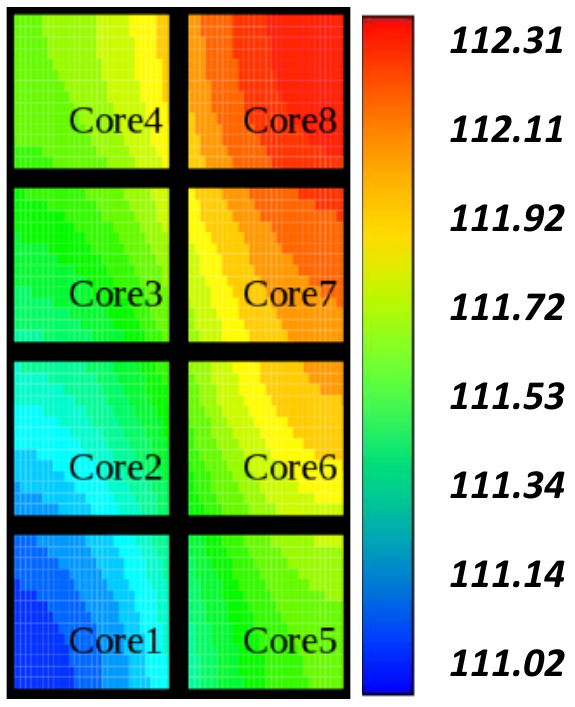}\label{fig:fig14b}}
        \hfil
        \subfloat[\cite{socci15}]{\includegraphics[width=0.28\textwidth]{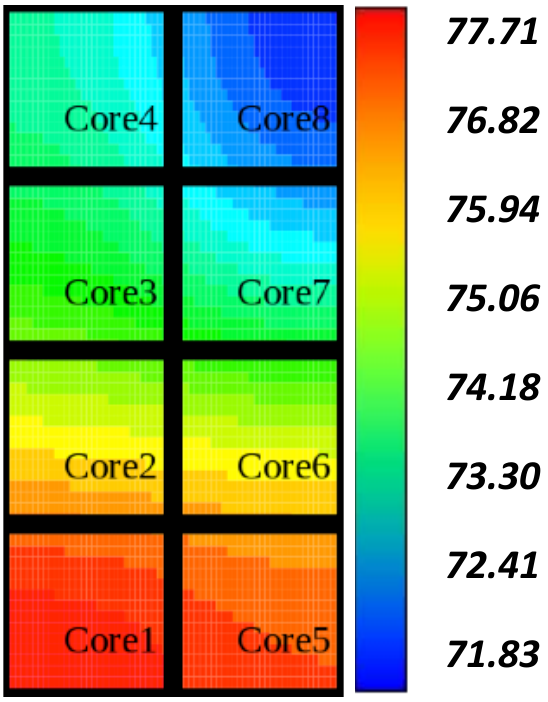}\label{fig:fig14c}}
		\caption{Thermal profiles of a random task graph example in different methods under worst-case scenario}
        \label{fig:expScenarios}
	\end{minipage}
\end{figure}

Now, we evaluate the power trace and thermal distribution of different methods of~\cite{socci15,Medina18,Ranjbar19} and our proposed method for a random task set example in the worst-case scenario, in terms of execution times and power consumption. It should be noted that the scale of temperature for each method is different in Fig.~\ref{fig:expScenarios}. Since we have generated many task graphs with different values of parameters (\textit{n}, \textit{d}, \textit{c} and \textit{U/c}), we choose one of the random task graphs with $d=10\%$, $n=50$, $c=8$ and $U/c=0.9$, which needs a high computational demand to show the results. Fig.~\ref{fig: Powerexp} shows the power traces, and Fig.~\ref{fig:expScenarios} shows the thermal distribution of the methods. As can be seen in Fig.~\ref{fig: Powerexp}, the peak power consumption is violated some times in the method of~\cite{Medina18,Ranjbar19}, and since the method, of~\cite{socci15} drops all LC tasks in the HI mode, which is not desirable, the task set finishes its execution earlier and also has less peak power consumption. Besides, Fig.~\ref{fig:expScenarios} depicts that the thermal distribution has not been managed in~\cite{Medina18,Ranjbar19}, while our approach reduces the hotspots and lowers the maximum temperature by 22.3$^\circ$C in this example.
Although the maximum temperature of~\cite{socci15} is lower than ours, the LC tasks’ QoS is zero since no LC tasks are executed in the HI mode.

\section{Power Trace of Real-Life Application Graph (CC) with Different Methods at Run-Time}
\label{A-CCPowTempavg}

\begin{figure}[t]
	\centering
	\includegraphics[width=0.99\columnwidth]{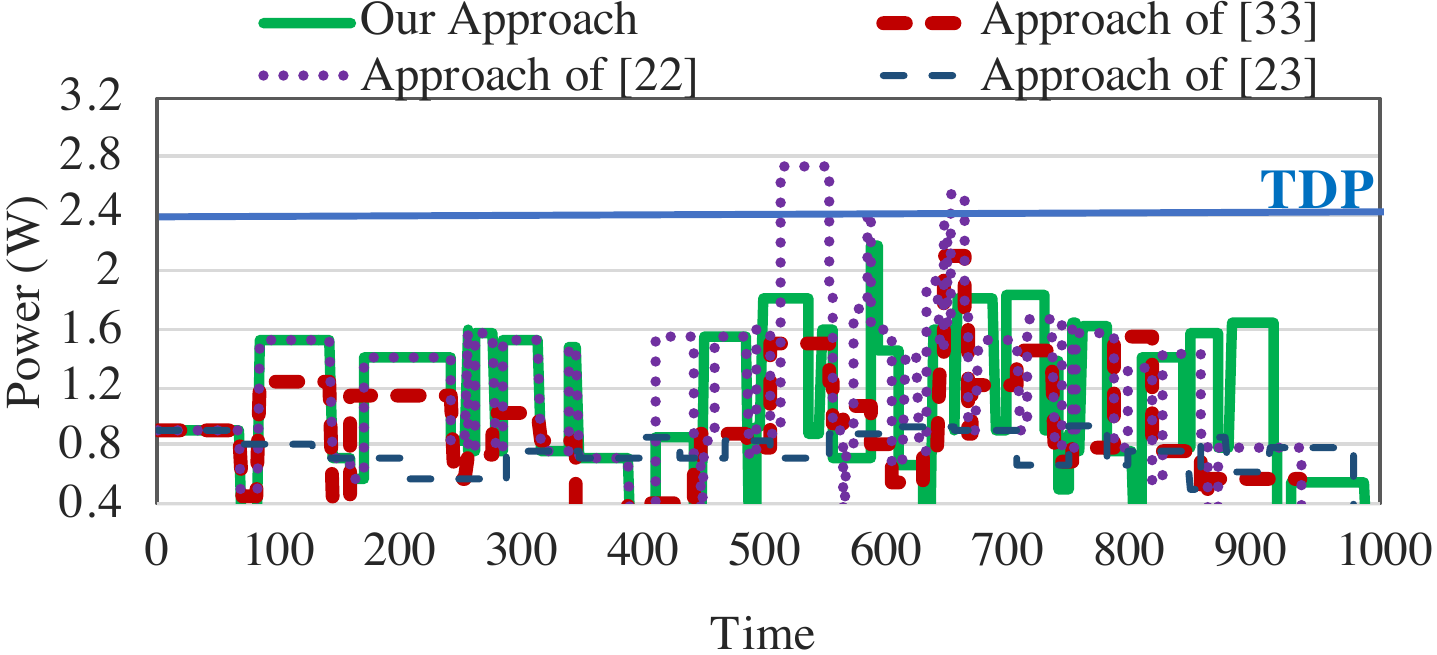}
	\caption{Run-time power trace of real-life task graph in different methods}
	\label{fig: PowerCCrun}
\end{figure}

Section~\ref{subsec:RuntimePowTrace} discussed the system behavior when the tasks finish their execution earlier than their worst-case execution time (WCET). The method of~\cite{Ranjbar19} uses the same task mapping and scheduling algorithm of~\cite{Medina18} at design-time and reduces the peak power by reclaiming the dynamic slack times at run-time (the difference between the WCET and actual execution time) and decreases the operating voltage and frequency level of cores while executing the tasks. We have analyzed our proposed method compared to different methods for an example in Section~\ref{subsec:RuntimePowTrace}. Since we have used a real-life task graph (Cruise Controller (CC)) to evaluate our proposed method, here we show the run-time system behavior for the real task graph in Fig.~\ref{fig: PowerCCrun}. As shown in this figure, the approach of~\cite{Medina18} still violates the power constraint (TDP), and since the approach of~\cite{Ranjbar19} has applied the DVFS technique, the peak power consumption has been reduced. However, 
the approach of~\cite{Ranjbar19} has degraded reliability due to the use of DVFS technique, which is not desirable for safety-critical applications. In addition, our proposed method could manage the peak power consumption in this real task graph, and also, since the method of~\cite{socci15} has dropped all LC tasks in the HI mode, the peak power consumption is also less than the TDP constraint.

\newpage

\xpatchcmd{\thebibliography}{%
  \usecounter{enumiv}%
}{%
  \usecounter{enumiv}%
  \setcounter{enumiv}{\value{mybibstartvalue}}%
}{}{}


\end{document}